\documentclass[a4paper,10pt]{article}
\usepackage{epsfig}
\renewenvironment{thebibliography}[1]
        {\begin{list}{\arabic{enumi}.}
        {\usecounter{enumi}\setlength{\parsep}{0pt}
         \setlength{\itemsep}{0pt} 
         \settowidth
        {\labelwidth}{#1.}\sloppy}}{\end{list}}
\begin{document}
\setcounter{tocdepth}{3}
\newcommand{\pb}{p\hspace*{-1.0ex}/}
\newcommand{\mat}[3]{\langle\, #1 \,|\, #2 \,|\, #3 \,\rangle}
\newcommand{\skal}[2]{\langle\, #1 \,|\, #2 \,\rangle}
\newcommand{\ket}[1]{|\, #1 \,\rangle}
\newcommand{\bra}[1]{\langle \, #1 \,|}
\newcommand{\olabbr}[1]{{\rm #1}}
\newcommand{\beq}{\begin{equation}}
\newcommand{\eeq}{\end{equation}}
\newcommand{\beqs}{\begin{displaymath}}
\newcommand{\eeqs}{\end{displaymath}}
\newcommand{\bea}{\begin{eqnarray}}
\newcommand{\eea}{\end{eqnarray}}
\newcommand{\beas}{\begin{eqnarray*}}
\newcommand{\eeas}{\end{eqnarray*}}
\newcommand{\vi}{{\bf x}_{\perp}}
\newcommand{\vj}{{\bf x}^{\prime}_{\perp}}
\date{\today}
\thispagestyle{empty}
\noindent\hspace*{\fill}  FAU-TP3-00/11 \\
\noindent\hspace*{\fill}  hep-ph/0010099 \\
\begin{center}\begin{Large}\begin{bf}  Compact Variables and Singular Fields in QCD  \\
\end{bf}\end{Large}\vspace{.75cm}
 \vspace{0.5cm} Frieder Lenz and Stefan W\"orlen\\ Institut f\"ur Theoretische Physik III \\
Universit\"at Erlangen-N\"urnberg, Erlangen Germany\\  
\end{center}
\vspace{1cm}

\date{\today}
\begin{abstract} \noindent 
Subject of our investigations is QCD formulated in terms of physical degrees of freedom. Starting from the Faddeev-Popov procedure, the canonical formulation of QCD is derived for static gauges. Particular emphasis is put on obstructions occurring when implementing gauge conditions and on the concomitant emergence of compact variables and singular fields. A detailed analysis of non-perturbative dynamics associated with such exceptional field configurations  within Coulomb- and axial gauge is described. We present evidence that compact variables generate confinement-like phenomena in both gauges and point out the deficiencies in achieving  a satisfactory non-perturbative treatment concerning all variables. Gauge fixed formulations are shown to constitute also a useful framework  for phenomenological studies. Phenomenological insights into the dynamics of Polyakov loops and monopoles in confined and deconfined phases are presented within axial gauge QCD.      
\end{abstract}

\newpage

\section{Introduction}

The existence of a common  formal structure in the theory of the fundamental interactions such as QED and QCD reflects the close similarity of the corresponding  fundamental processes at high momenta. This common formal structure remains manifest in perturbation theory.  The high precision predictions of perturbative QED have found their  counterpart in the quantitative  verification of perturbative QCD predictions at large momenta. The diversity of low energy phenomena on the other hand  is a much less obvious consequence of the dynamics of gauge theories. While the low energy properties of the weak interaction are understood in terms of the  Higgs mechanism as the basic non-perturbative element, emergence of the characteristic  low energy phenomena of QCD, in particular of confinement, remains a central topic in the analysis of gauge theories. Unlike the Higgs mechanism, confinement is an intrinsic property of Yang-Mills theories. Identification of the seeds of confinement and its consequences  is not possible within the realm of perturbation theory. 

Unity of  formal structures in the theory of fundamental interactions and diversity of low energy phenomena are compatible due to the redundancy of variables - an intrinsic property of gauge theories.  In turn, essential differences in the structure of
QED and QCD can be expected to become manifest when formulating  gauge theories in terms of physical, unconstrained variables.
Elimination of redundant variables in QCD necessarily introduces non-perturbative elements. In general, the process of  the gauge fixing  cannot be carried out completely. In QED it is possible to impose a global -- for all field configurations valid -- gauge condition and thereby to eliminate all the redundant variables. It is known~\cite{SING78} that in QCD,  a large class of gauge conditions  leads to gauge ambiguities i.e.\ there exist gauge equivalent field configurations which satisfy the imposed gauge condition. Elimination of variables  has to respect the complex structure of the manifold of gauge orbits. Depending on the gauge condition, this might be achieved by restricting the domain of definition of the physical variables or at  the expense of introducing coordinate singularities and thus of
including singular gauge field configurations in  gauge fixed formulations. It is tempting to associate with this basic geometrical difference between QED and QCD the characteristic phenomenological differences between the strong and electromagnetic forces at low energies. Formation of Gribov horizons~\cite{GRIB78} or condensation of magnetic
monopoles~\cite{MAND80,THOO81} represent two prominent proposals  in which the
emergence of confinement has been  linked directly to obstructions when enforcing a global gauge condition. 

We will discuss in this work  progress and problems in the study of gauge fixed formulations and present an analysis of non-perturbative properties of QCD in Coulomb- and axial gauge. In these gauges it is possible to eliminate completely redundant variables and to formulate thereby the theory in terms of unconstrained variables without need for introducing  gauge fixing terms or ghosts. These ``static'' gauge choices neither involve the time component of the gauge field nor derivatives with respect to time and therefore a standard Hamiltonian description of the dynamics with positive norm states can be developed. Necessarily, manifest Lorentz-invariance is lost in these gauges; in axial gauge  rotational invariance is not manifest either. We start these studies with a description of the Faddeev-Popov procedure in the quantization of gauge theories and proceed to the canonical formulation  with an expression for the Hamiltonian of QCD in static gauges as our main result. Our discussion will emphasize from the beginning   the role of possible obstructions in carrying out this procedure. After implementation of the gauge condition, characteristic properties become manifest which distinguish for instance  the canonical formulations of QCD and QED. While the corresponding Hamiltonians both  contain ``centrifugal'' energies arising, as in mechanical systems,  from elimination of kinetic terms (electric fields) only in QCD these terms are intrinsically dynamical. With these results, a  framework will have been established which makes the dynamics  accessible to approximative treatments. In such formulations no difficult  constraints imposed by local gauge invariance such as Ward-identities have to be observed. This is essential for our dynamical studies of QCD in  Coulomb- and axial gauge. In general we concentrate our studies on pure Yang-Mills theories. Inclusion of matter fields  leaves the  structure of the gauge fixed theory  essentially unchanged in  Coulomb- or axial gauge. This is different if the gauge condition depends on matter fields or is exclusively given in terms of these, as in the unitary gauge of the Georgi-Glashow model.  The unitary gauge condition gives rise to obstructions in the gauge fixing procedure closely resembling those encountered in the axial gauge - or more generally in diagonalization gauges. We include in this formal section a short discussion of this model.    

Representation of QCD in  Coulomb gauge  has played a very important role in the development of gauge fixed formulations~\cite{BJOR,CHLE,GESA,CREU}.
In QED,  the Coulomb-gauge  is singled out as the gauge
in which static charges do not radiate. In QCD this is not the case, gluons and color spin remain coupled irrespective of the gauge choice. The focus of our discussion of QCD in Coulomb-gauge will be the role of Gribov horizons, i.e.\ the consequences of restrictions in the range of integration over gauge field variables.  Gribov horizons are invisible in perturbation theory, they  however limit  large amplitude oscillations. We nonetheless will start this discussion with a perturbative calculation and demonstrate how asymptotic freedom will emerge in a formulation where only physical variables are present. It will be seen that polarization mechanisms are present which involve the instantaneous Coulomb interaction and which are induced by the centrifugal terms in the Hamiltonian. They cannot be associated with the coupling to intermediate physical states and therefore do not necessarily generate screening. These very same polarization mechanisms will subsequently be shown to be severely affected by the presence of a Gribov horizon. We will describe Gribov's approach in dealing approximatively with the compactness of the gauge fields and we will  discuss the dynamical implications. In particular we will study the interaction energy of static color charges and indicate the possible effects of the compactness on the structure of the vacuum.  

Formulation of QCD in the axial gauge requires a slightly more complicated setting. For a proper definition of this gauge, space has to be compact in at least one direction, otherwise this formulation will be plagued by infrared singularities. The axial gauge condition eliminates the component of the gauge field corresponding to the compact direction  up to the eigenvalues of the Polyakov loops  winding around the finite spatial extension. In axial gauge we will encounter compact variables and singular fields. The restriction in the range of integration affects special variables, the eigenvalues of the Polyakov loops. The corresponding Gribov horizon has a simple geometrical origin and is known explicitly. The formalism suggests to treat these variables in analogy to quantum mechanical particles enclosed in an infinite square well. We will describe this procedure and survey the dynamical implications which are far reaching, since these variables, although only a small subset of the dynamical variables, serve as order parameter-fields for the phases of non-Abelian Yang-Mills theories. Implementation of the axial gauge condition requires a diagonalization of the Polyakov loops which in turn gives rise to singular fields. Their structure will be described and, in absence of a viable scheme for treating such variables,  arguments will be presented concerning their relevance for the dynamics of QCD.

\section{Path Integral and Canonical Formulation}

\subsection{Yang-Mills Fields}

Central to the following studies is the discussion of the elimination of the redundant variables. We start with a brief description of the Faddeev-Popov procedure and will establish for a certain class of gauge conditions the connection to the canonical formalism. In  first applications we discuss the structure of  QCD and QED in the Coulomb-gauge and  the canonical structure of the Georgi-Glashow model in the unitary gauge.\\ 
We use the following notation and conventions. Gauge fields and their color components are related by
$$A_{\mu}(x)=A_{\mu}^{a}(x)\frac{\lambda^{a}}{2},$$
where the color sum runs over the $N^{2}-1$ generators of SU$(N)$.   
The covariant derivative, field strength tensor and its color components are defined by 
\beqs
D_{\mu} = \partial_{\mu}+igA_{\mu}, 
\eeqs
\beqs
F^{\mu\nu} = \frac{1}{ig}[D_{\mu},D_{\nu}],\quad F_{\mu\nu}^a = \partial_\mu A_\nu^a - \partial_\nu A_\mu^a 
             - gf^{abc} A_\mu^b A_\nu^c .
\eeqs
Action and Lagrangian are
$$ S[A] = \int d^{4} x {\cal L} =  - \frac{1}{4}\int d^{4} x\,\, F^{\mu\nu a} F_{\mu\nu}^a = - \frac{1}{2}\int d^{4} x\,\mbox{tr}\, F^{\mu\nu} F_{\mu\nu}.$$    
As indicated by the notation,  in general,  color labels will be suppressed. For the canonical formulation of QCD one introduces chromo- electric and magnetic fields which, as in QED, are related to the field strength tensor by 
\beqs
E^{i a}\left(x\right)=-F^{0i a}\left(x\right) \quad , \quad  
B^{i a}\left(x\right)=-\frac{1}{2}\epsilon^{ijk}F^{jk a}\left(x\right) \ .
\eeqs
Gauge transformations and transformed fields are written as
\beq
A_{\mu} \left(x\right) \rightarrow A^{\, \left[\,\Omega\right] }_{\mu} \left(x\right)= \Omega \left(x\right) \left(A_{\mu}\left(x\right)+\frac{1}{ig} \partial _{\mu} \right) \Omega^{\dagger} \left(x\right) \ ,
\label{I13}
\eeq
where the gauge transformations are parametrized by  gauge functions $\alpha(x)$ 
\beqs
\Omega \left(x\right)= \exp\left\{ig \alpha \left(x\right)\right\}=\exp\left\{ig \alpha^{a} \left(x\right) \frac{\lambda ^{a}}{2}\right\} \ ,
\eeqs
which for gauge transformations close to the identity yields
\beq
A_{\mu} \left(x\right) \rightarrow  A_{\mu}^{\,\left[\,\alpha\right]} \left(x\right)  = A_{\mu} \left(x\right)- D _{\mu}\alpha \left(x\right) -\frac{ig}{2}[\alpha,D_{\mu}\alpha]  .
\label{I16}
\eeq
In path integral quantization of gauge theories redundant variables are eliminated by imposing a gauge condition
$$ f[A]=0 , $$
which is supposed to eliminate all the gauge copies of a certain field configuration $A$. 
In other words, the functional $f$ has to be chosen such that  for a given field configuration the equation
$$ f[ A^{\, \left[\,\Omega\right]}\,]=0$$ 
determines uniquely the gauge transformation $\Omega$. If successful, the set of all gauge equivalent fields, the gauge orbit, is represented by exactly one representative and the resulting generating functional 
\beq
Z \left[J\right]= \int d\left[A\right]\Delta_{f}\left[A\right]
\delta\left(f\left[A\right]\,\right) e^{i S \left[A\right]+i \int d^{4}x J^{\mu}  A_{\mu}}
\label{IIb15} 
\eeq
is given as a sum over these gauge orbits. The Faddeev-Popov determinant
\bea
\Delta_{f}\left[A\right] = |\det M \ |\  , \ \  M = \frac{\delta f[A^{\,[\alpha]}]}{\delta\alpha}\Big|_{\alpha = 0} && \label{FPD}
\\
M\left(x,y;a,b\right) =  \int d^{4} z\, m^{\mu}\left(x,z;a,c\right)D_{\mu}^{cb}(z)\delta(z-y) &,&
m^{\mu}\left(x,y;a,b\right) = \frac{\delta  f^{a}_{x} \left[A\right]}{\delta A^{b}_{\mu} \left(y\right)}, \nonumber
\eea
ensures that the weight of a gauge orbit in the integral is independent of the representative chosen, i.e.\ independent of the gauge function $\alpha(x)$ (cf.\ Eq.(\ref{I16})). If $ \Delta_{f}\left[A\right]$ vanishes, the gauge condition exhibits a quadratic or higher order zero. This implies that at this point in function space, the gauge condition is satisfied by at least two gauge equivalent configurations. Vanishing of $ \Delta_{f}\left[A\right]$ implies the existence of zero modes associated with $M$ 
\begin{equation}
\label{d_chi0}
M \chi_{0} = 0
\end{equation}
and therefore the gauge choice is ambiguous. 
The (connected) spaces of gauge fields which make the gauge choice ambiguous
$${\cal M}_{H} = \left\{A \big|\,\det M = 0 \right\} $$   
are called  Gribov horizons. Around Gribov horizons, pairs of  infinitesimally close gauge equivalent fields exist which satisfy the gauge condition. For a field  $A_{\mu}$  close to the Gribov horizon we write
$$A_{\mu}= A_{\mu}^{H}+a_{\mu} ,\quad  M[A^{H}]\chi_{0} = 0, $$
and treat $a_{\mu}$ as a small quantity. A Gribov copy 
$$A_{\mu}^{\prime}= A_{\mu}^{H}+a_{\mu}^{\prime}$$
must be connected to $A_{\mu}$ by a gauge transformation. The gauge function $\alpha$ must therefore satisfy (cf.\ Eq.(\ref{I16}))
$$0=f[A^{\prime}=A^{[\alpha]}\,]-f[A]\approx -m^{\mu}( D _{\mu}\alpha \left(x\right) +\frac{ig}{2}[\alpha,D_{\mu}\alpha] ) .$$
For this discussion the gauge condition has been assumed to be linear in the gauge fields.
With the  Ansatz for the gauge function 
$$\alpha = \chi_{0} + \delta \chi , \quad \delta \chi \ll \chi_{0}, $$
with arbitrarily normalized $\chi_{0}$ and $\delta \chi$ orthogonal to $\chi_{0}$, the above condition can be rewritten as
$$m^{\mu}D_{\mu}\delta \chi = -ig([\,a_{\mu},\chi_{0}]+\frac{1}{2}\,[\chi_{0},D_{\mu}\chi_{0}]) \quad 
\mbox{with} \quad
D = D[A^{H}].$$
This condition can be solved for $\delta \chi$ provided the right hand side of the equation  has a vanishing projection on $\chi_{0}$. The requirement
$$(\chi_{0},[a_{\mu},\chi_{0}])+(\chi_{0},\frac{1}{2}[\chi_{0},D_{\mu}\chi_{0}]) = 0$$
determines the normalization of $\chi_{0}.$ 
Thus the Gribov copy to $A_{\mu}$ has been constructed 
$$a_{\mu}^{\prime}= a_{\mu} -D_{\mu}\chi_{0}.$$
Perturbation theory in $a^{(\prime)}_{\mu}$ yields the eigenvalues $\lambda$ of of the Faddeev-Popov operators corresponding to this pair of gauge equivalent fields 
$$ \lambda[A^{(\prime)}] (\chi_{0},\chi_{0}) =  (\chi_{0},m^{\mu} D_{\mu}[A^{(\prime)}]\chi_{0}) \approx igm^{\mu}(\chi_{0},[a_{\mu}^{(\prime)},\chi_{0}]).$$
Using the above normalization condition shows the two eigenvalues to be of opposite sign
$$ \lambda[A^\prime]=- \lambda[A]\,. $$
 The spectrum of the Faddeev-Popov operator corresponding to the gauge field beyond the Gribov horizon possesses a bound state.\\ 
If on the other hand two gauge fields satisfy the gauge condition and are separated by an infinitesimal gauge transformation $\epsilon\alpha(x)$, then 
$$M\alpha(x) =0 $$
and the two fields are separated by a Gribov horizon.
  The region beyond the horizon thus contains gauge copies of fields inside the horizon. In general one therefore needs additional criteria to select exactly one representative of the gauge orbits. The structure of Gribov horizons and of the space of fields which contain no Gribov copies depends on the choice of the gauge. At this point we do not specify the procedure further  but rather associate an infinite potential energy ${\cal V}[A]$ with every gauge copy of a configuration which already has been taken into account, i.e.\ after gauge fixing, the action is supposed to contain implicitly this potential energy
\beq
\label{VA}
S[A] \rightarrow S[A] -\int d^4 x\,{\cal V}[A] . 
\eeq
The above expression for the generating functional can serve as a starting point for deriving the Hamiltonian of canonical quantization. To this end we  restrict the class of gauge conditions by assuming that $f$ is independent of the time component of $A$ and does not contain time derivatives of the gauge fields, 
\beq
\label{statgc}
 f[{\bf A}] = 0. 
\eeq
We now convert the functional integral in Eq.(\ref{IIb15}) into a phase space functional integral. We write the electric  field components  as
$$F_{0i}^a = \partial_0A_i^a - (D_iA_0)^a$$
and  introduce auxiliary electric field variables to eliminate terms quadratic in $A_{0}$. In the resulting expression  
\beqs
Z[J] =
\int\! d[{\bf E}]\,d[A_\mu]\,\delta(f[{\bf A}])\,
|\det({\bf D} \frac{\delta f}{\delta {\bf A}})|
 \cdot \exp\left \{ i\int\!d^4\!x
\left ( -\frac{1}{2}\,({\bf E}^2+{\bf B}^2) 
- {\bf E}\cdot(\dot{{\bf A}} - {\bf D} A_0) + J^{\mu}A_{\mu}\right )
    \right \}
\eeqs
$A_{0}$ can be integrated out
\beqs
Z[J] = 
\int\! d[{\bf E},{\bf A}]\,
\delta(f[{\bf A}])\,\,|\det({\bf D} \frac{\delta f}{\delta {\bf A}})|\,\,\delta({\bf D}\cdot{\bf E} - J^{0})
\cdot\exp\left \{ i\int\!d^4\!x
\left (-{\bf E}\cdot\dot{{\bf A}} -\frac{1}{2}\,({\bf E}^2+{\bf B}^2) -{\bf J}{\bf A}\right )
    \right \}.
\eeqs
In this procedure, the Gauss law appears as a constraint which will be used to eliminate the electric field variables which are conjugate to the gauge fields eliminated by the gauge condition $f[{\bf A}]$. To identify the appropriate electric field variables we decompose the function space ${\cal M}_{{\bf E}}$ of electric fields as
\beq
\label{Mtr}
 {\cal M}_{{\bf E}} = {\cal M}_{\perp}\otimes {\cal M}_{\parallel} 
\eeq
with kernel ${\cal M}_{\perp}$ and quotient space ${\cal M}_{\parallel}$
$${\cal M}_{\perp} = \left\{{\bf E}_{\perp}\,\Big|\,\frac{\delta f}{\delta {\bf A}}{\bf E}_{\perp}=0\right\},\quad {\cal M}_{\parallel} = {\cal M}_{{\bf E}}/{\cal M}_{\perp}.$$  
Assuming $\delta f/\delta A$ to be regular (cf.\ Eq.(\ref{FPD})), i.e.
$$\det m_{\mu}^{\dagger}m^{\mu} \neq 0 ,$$ 
we can define the projector on ${\cal M}_{\parallel} $ ($m =(0,{\bf m}))$
$$P_{\,\parallel} = {\bf m}^{\dagger}\frac{1}{({\bf m}{\bf m}^{\dagger})}  {\bf m}$$
and can represent any $ {\bf E}_{\parallel}$
\beq
\label{defg}
  {\bf E}_{\parallel} = {\bf m} ^{\dagger}\,\varphi .
\eeq
For fixed gauge field ${\bf A}$, the functional integral over the electric field variables can - up to a field dependent normalization $N[{\bf A}] $ - be written  in terms of integrations over ${\bf E}_{\perp}$ and the scalar field $\varphi$
$$\int d[{\bf E}]= N[{\bf A}]\int d[{\bf E}_{\perp}] d[{\bf E}_{\parallel}]\int d[\varphi]\,\delta({\bf E}_{\parallel} - {\bf m}^{\dagger} \varphi).$$ 
With the 
 definition
$$ \int d[{\bf A}] \delta(f[{\bf A}]) N[{\bf A}]= \int d[{\bf A}_{\perp}]\tilde{N}[A] ,$$
the final result for the phase space integral 
\bea
Z[J] &=& 
\int\! d[{\bf E}_{\perp},{\bf A}_{\perp}]\,\cdot\exp i\int\!d^4\!x
\Big\{ -{\bf E}_{\perp}\cdot\dot{{\bf A}}_{\perp} -\frac{1}{2}\,\left({\bf E}_{\perp}^2+{\bf B}[{\bf A}_{\perp}]^2\right)
\nonumber\\
& & - \frac{1}{2} \Big( {\bf m}^{\dagger}\,\frac{1}{{\bf D}{\bf m}^{\dagger}}({\bf D}{\bf E}_{\perp}-J^{0})\Big)^{2} -{\bf J}_{\perp}{\bf A}_{\perp} \Big\} +\ln \tilde{N}[{\bf A}_{\perp}].
\label{PSI2}
\eea
is obtained.
This expression identifies $-{\bf E}_{\perp}$ and ${\bf A}_{\perp}$ as conjugate variables and defines the Hamiltonian in the presence of external color currents 
\beq
\label{ham}
H = \int d^{3}x \Big\{ \frac{1}{2} \left({\bf E}_{\perp}^2+{\bf B}[{\bf A}_{\perp}]^2\right) + \frac{1}{2} \Big({\bf m}^{\dagger}\,\frac{1}{{\bf D}{\bf m}^{\dagger}}({\bf D}{\bf E}_{\perp}-J^{0}\,)\Big)^{2}
+{\bf J}_{\perp}{\bf A}_{\perp} \Big\}  +i\ln \tilde{N}[{\bf A}_{\perp}].
\eeq
This Hamiltonian contains  the energy density $\frac{1}{2} ({\bf E}_{\perp}^2+{\bf B}_{\perp}^{2})$ of the physical, unconstrained variables. The additional term in the Hamiltonian is  the kinetic energy of the eliminated variables. It appears in exactly the same way as the centrifugal energy when eliminating angular variables in spherically symmetric problems of quantum mechanics. As the centrifugal barrier, this additional term always generates repulsion.  In the quantum mechanical problem, infinite repulsion is obtained  at the center where the angles are ill defined and, in gauge theories, where the gauge condition is ambiguous, i.e.\ where with ${\bf D}{\bf m}^{\dagger}$ the Faddeev-Popov operator $M$ possesses one or more zero modes. In both cases, the configurations where the centrifugal term diverges describe the limits in the range of definition of the corresponding dynamical variables. Integration beyond these limits is effectively eliminated by the potential ${\cal V}[A]$ included in the action (cf.\ Eq.(\ref{VA})).  In quantum mechanics this procedure  corresponds to implement  the boundary condition at $r=0$ by the action of an infinite square well $V_{0}\theta(r)$. With this interpretation, it becomes plausible that phase space path-integral and  Hamiltonian do not contain the Faddeev-Popov determinant which, in the generating functional, suppresses contributions from fields on the Gribov horizon. With the action, also the  Hamiltonian contains  ${\cal V}[A]$ and we therefore conclude that the Hamiltonian acts in the space of reduced wavefunctionals $\hat{\psi}[A]$. In particular  these wave-functionals have to vanish along a Gribov horizon 
$$ \hat{\psi}[A]\,\big|_{A\in {\cal M}_{H}} = 0 , $$
which separates infinitesimally close Gribov copies.              

The paradigm in the canonical description of gauge theories is provided by QED in the Coulomb-gauge and  in a first application of the general procedure we discuss the formal structure of Coulomb-gauge QCD and QED. 
In Coulomb-gauge  
\beq
\label{clbg}
f[A] = \mbox{div}\, {\bf A}
\eeq
one eliminates the longitudinal gauge fields. With
$$\frac{\delta f^{a}_{x}[A]}{\delta A_{i}^{b}(y)} = \,\delta^{ab}\partial_{i}\,\delta(x-y),\quad  {\bf m}^{\dagger} = \mbox{\boldmath$\nabla $} ,\quad \Delta_{f}[A]= |\det {\bf D}\mbox{\boldmath$\nabla $}\,|$$
the following expression for the generating functional
\beqs
Z \left[J\right]= \int d\left[A\right]\,|\det {\bf D}\mbox{\boldmath$\nabla $}\,|\,  \delta\left( \mbox{div}\, {\bf A}\right)\, e^{i S \left[A\right]+i \int d^{4}x J^{\mu}  A_{\mu}}.  
\eeqs
is obtained; in this and other gauges discussed below the normalization $\tilde{N}[{\bf A}]$ (cf.\ Eq.(\ref{PSI2})) is field independent and has been dropped.
The transition to the canonical formalism is conceptually and technically simple for  gauge conditions linear in $A_{\mu}$ such as (\ref{clbg}). In Coulomb-gauge ${\cal M}_{\perp}$ and ${\cal M}_{\parallel}$ are the space of transverse and longitudinal vectorfields respectively 
$${\cal M}_{\perp} = \left\{{\bf E}_{\mbox{tr}}\right\}= \left\{{\bf E}\,\Big|\ \mbox{div}\,{\bf E}=0 \right\},\quad {\cal M}_{\parallel} = \left\{{\bf E}\,\Big|\ \mbox{{\bf rot}}\,{\bf E}=0 \right\}$$
From Eq.(\ref{ham}) the Coulomb-gauge Hamiltonian is easily derived
\beq
\label{cgha}
H = \int d^{3}x \Big\{ \frac{1}{2} {\bf E}_{\mbox{tr}}^2+{\bf B}[{\bf A}_{\mbox{tr}}]^2
- \frac{1}{2} ({\bf D}\,{\bf E}_{\mbox{tr}}-J^{0}\,)\frac{1}{{\bf D}\,\mbox{\boldmath$\nabla $}}\,\Delta \,\frac{1}{{\bf D}\,\mbox{\boldmath$\nabla $}}({\bf D}\,{\bf E}_{\mbox{tr}}-J^{0}\,)+ {\bf J}_{\perp}{\bf A}_{\perp}\Big\} .
\eeq
The corresponding expression for the Maxwell theory is obtained by replacing ${\bf D} \rightarrow \mbox{\boldmath$\nabla $}$
\beqs
 H_{M} = \int d^{3}x \left\{ \frac{1}{2} \left({\bf E}_{\mbox{tr}}^2+{\bf B}[{\bf A}_{\mbox{tr}}]^2 -J^{0}\frac{1}{\Delta}J^{0} \right) + {\bf J}_{\perp}{\bf A}_{\perp}\right\}, \quad {\bf B} =  \mbox{\bf curl}\, {\bf A}_{\perp}.  
\eeqs
In Coulomb-gauge QED, static charges do not couple to the radiation field. This is a very specific property of  Coulomb-gauge electrodynamics. In axial gauge QED, for instance, static charges couple to the radiation field although energy  conservation prohibits radiation to result from such a coupling. In non-Abelian theories coupling of static color charges to the dynamical degrees of freedom cannot be eliminated by the gauge choice. The color spin of static charges always gives rise to a non-trivial coupling to the ``radiation field''. In Coulomb-gauge QED, the $1/r$ is read off directly from the Hamiltonian. As in the general case, the Coulomb-energy appears as a centrifugal barrier, however it is independent of dynamical degrees of freedom.
 
\subsection{Matter Fields}

In this concluding part of our formal considerations we describe the   modifications which become necessary when  matter is coupled to the gauge fields. For static gauge conditions of the type (\ref{statgc}) the modifications are technical. For the important case of  fermions in the fundamental representation as described by the additional Lagrangian 
$${\cal L}_{f} = \bar{\psi}(i\gamma^{\mu}D_{\mu}-m)\psi$$
the modifications  can be easily derived with the help of the substitution 
$$J_{\mu} \rightarrow J_{\mu} -g\bar{\psi}\gamma_{\mu}\psi $$
leading to the following form of the Hamiltonian
\bea
\label{quha}
H &=& \int d^{3}x \Big\{ \frac{1}{2} \left({\bf E}_{\perp}^2+{\bf B}[{\bf A}_{\perp}]^2\right) + \frac{1}{2} \left({\bf m}^{\dagger}\,\frac{1}{{\bf D}{\bf m}^{\dagger}}({\bf D}{\bf E}_{\perp}-J^{0}-g\psi^{\dagger}\psi)\right)^{2} 
\nonumber \\
&& +\psi^{\dagger}\mbox{\boldmath$\alpha$}(\frac{1}{i} \mbox{\boldmath$\nabla$}-g{\bf A})\psi +m \psi^{\dagger}\beta\psi +{\bf J}_{\perp}{\bf A}_{\perp}\Big\} .
\eea
Since the gauge condition has not been affected by the presence of the fermions, the structure of the gauge fixed theory is not significantly altered. Charged matter provides sources of the fields which add to the external sources  The same remarks apply for the case of the Georgi-Glashow model~\cite{Georgi72} when using a static gauge condition.
 In the Georgi-Glashow model bosons in the adjoint representation of SU$(2)$ are coupled to SU$(2)$ Yang-Mills fields. The Lagrangian is
\beqs
{\cal L}_{GG}= \frac{1}{2}D_{\mu}\phi D^{\mu}\phi -V(\phi^{2}).
\eeqs
Under a gauge transformation $\Omega$ (cf.\ Eqs.(\ref{I13},\ref{I16})) the matter field transforms covariantly
\beq
\label{adjgt}
\phi \left(x\right) \rightarrow \phi^{\, \left[\,\Omega\right] } \left(x\right)= \Omega \left(x\right) \phi\left(x\right) \Omega^{\dagger} \left(x\right) \ .
\eeq
 In this case it is of interest to generalize the formalism and allow for dependencies of the gauge condition on the matter fields $\phi$
$$f= f[{\bf A},\phi].$$
We proceed as above from the generating functional 
\beqs
Z \left[J\right]= \int d\left[A,\phi\right]\Delta_{f}[{\bf A},\phi]
\,\,\delta\left(f\left[{\bf A},\phi\right]\,\right) e^{i S \left[A,\phi\right]+i \int d^{4}x J^{\mu}  A_{\mu}},
\eeqs
with
$$\Delta_{f}[{\bf A},\phi] = \Big|\det \frac{\delta f[{\bf A}^{[\alpha]},\phi^{[\alpha]}]}{\delta \alpha}\big|_{\alpha=0}\Big| \, .$$
We derive the phase space form of $Z[J]$ by introducing fields $\pi(x)$ conjugate to the boson fields $\phi$ in addition to the electric fields ${\bf E}(x)$. 
In this way the time component $A_{0}$ appears again linearly in the exponent of the integrand and after integration yields the Gau\ss\ law constraint
\beas
&& Z[J] = 
\int\! d[{\bf E},{\bf A},\pi,\phi]\,\Delta_{f}[{\bf A},\phi]
\delta(f[{\bf A},\phi])\,\,\,\delta({\bf D}\cdot{\bf E} - J^{0}-ig[\pi,\phi])\nonumber\\
&& \exp\left \{ i\int\!d^4\!x
-{\bf E}\cdot\dot{{\bf A}}-\pi\dot{\phi} -\frac{1}{2}\,({\bf E}^2+{\bf B}^2+\pi^{2}+({\bf D}\phi)^{2}+V(\phi^{2})) -{\bf J}{\bf A}
    \right \}.
\eeas
As in the above case of fundamental fermions, if matter independent gauge conditions are used 
$$f[{\bf A},\phi]=f[{\bf A}] \, ,$$
only minor changes occur and the result can be read off by inspection (cf.\ Eq.(\ref{PSI2}))
\beas
H &=& \int d^{3}x \Big\{ \frac{1}{2} \left({\bf E}_{\perp}^2+{\bf B}[{\bf A}_{\perp}]^2+\pi^{2}+({\bf D}\phi)^{2}+V(\phi^{2})\right)
\nonumber \\
&&  + \frac{1}{2} \left({\bf m}^{\dagger}\,\frac{1}{{\bf D}{\bf m}^{\dagger}}({\bf D}{\bf E}_{\perp}-ig[\pi,\phi]-J^{0}\,)\right)^{2} +{\bf J}_{\perp}{\bf A}_{\perp}\Big\} .
\eeas
For the Georgi-Glashow model  the  option of matter dependent gauge conditions is relevant  for displaying the physics content of the theory.  In particular  the transformation property (\ref{adjgt}) suggests to use the gauge freedom to diagonalize the matter field. It is convenient to implement this condition with the help of an auxiliary field $\rho(x)$. We write the gauge condition as
\beq
\label{GGgc}
   f[{\bf A},\phi] =  f[\phi]=\phi - \rho\frac{\tau_{3}}{2}+\tilde{f}[{\bf A}^{3}],\quad \rho(x)\ge 0
\eeq
and integrate in $Z[J]$ over $\rho$. The diagonalization condition only would not constitute a complete gauge fixing; without $\tilde{f}[{\bf A}^{3}]$ the gauge condition (\ref{GGgc}) is invariant under Abelian gauge transformations 
$$\Omega(x) = e^{ig\alpha^{3}(x)\tau_{3}/2} .$$
This U(1) gauge symmetry has to be removed by the  subsidiary gauge condition
$ \tilde{f}[{\bf A}^{3}]$. The Faddeev-Popov determinant evaluated for fields which satisfy the gauge condition is easily calculated
$$\frac{\delta f^{a}_{x}[{\bf A}^{[\alpha]},\phi^{[\alpha]}]}{\delta \alpha^{b}_{y}}\big|_{\alpha=0}  = -g\epsilon^{ab3}\delta(x-y)\rho(x)+\frac{\delta \tilde{f}^3_{x}[{\bf A}^{[\alpha]\, 3 }]}{\delta \alpha^3_{y}}\big|_{\alpha=0}\delta^{a3}\delta^{b3} $$
and the following expression for the generating functional is obtained
$$Z \left[J\right]= \int d\left[A,\phi\right]\,D[\rho]\Delta_{\tilde{f}}[{\bf A}^{3}]
\,\,\delta\left(\tilde{f}\left[{\bf A}^{3}\right]\,\right) e^{i S \left[A,\rho\tau_{3}/2\right]+i \int d^{4}x J^{\mu}  A_{\mu}}.$$
The Faddeev-Popov determinant obviously gives rise to the non-trivial measure
\beq
\label{rhomeas}
  D[\rho] =\prod_{x}\rho^{2}(x)g^{2} d\rho(x).
\eeq
The transition to the phase space integral can now be performed with the result
\beas
Z[J] &=& 
\int\! D[{\bf E},{\bf A},\pi_{\rho},\rho]\,\Delta_{\tilde{f}}[{\bf A}^{3}]
\,\,\delta\left(\tilde{f}\left[{\bf A}^{3}\right]\,\right)\,\,\,\delta(({\bf D}\cdot{\bf E})^{3} - J^{0,3})\cdot\exp \Big\{ i\int\!d^4\!x
\Big( -{\bf E}\cdot\dot{{\bf A}}-\pi_{\rho}\dot{\rho}
\nonumber\\
& & -\frac{1}{2}\,\big[{\bf E}^2+{\bf B}^2+\pi_{\rho}^{2}+({\bf D}\,\rho\, \frac{\tau_{3}}{2} )^{2}+V(\rho^{2})
+ \frac{1}{g^{2}\rho^{2}}\sum_{a=1,2}(({\bf D}{\bf E})^{a}-J^{a,0})^{2} \big] - {\bf J}{\bf A} \Big) \Big \}.
\eeas 
 After choosing $\tilde{f}$ the Hamiltonian is easily obtained from this expression. Imposing the (Abelian) Coulomb-gauge condition
$$\tilde{f}\left[{\bf A}^{3}\right] = \mbox{div}{\bf A}^{3} $$
straightforward application of the above  procedure yields
\beas
H &=& \int d^{3}x \Big\{ \frac{1}{2} \big( {\bf E}_{\perp}^2+{\bf B}[{\bf A}_{\perp}]^2+\pi_{\rho}^{2}   +V(\rho^{2})  +   \frac{1}{g^{2}\rho^{2}}\sum_{a=1,2}(({\bf D}{\bf E})^{a}-J^{a,0})^{2}  \nonumber \\
& & + ({\bf D}\,\rho\, \frac{\tau_{3}}{2} )^{2} -\left(J^{0}+ig[{\bf E}_{\perp},{\bf A}_{\perp}]\right)^{3} \frac{1}{\Delta}\left(J^{0}+ig[{\bf E}_{\perp},{\bf A}_{\perp}]\right)^{3} \big) +{\bf J}_{\perp}{\bf A}_{\perp}\Big\} .
\eeas
The dynamical variables appearing in the Hamiltonian are  the pair of (one component) scalar fields $\rho,\pi_{\rho}$ and the pair of color vectorfields ${\bf A}_{\perp},{\bf E}_{\perp}$ whose longitudinal color 3-components vanish.
These expressions illustrate our general remarks concerning the structure of gauge fixed theories. In the process of elimination of 2 components of the scalar fields and of the longitudinal color 3-component gauge field two centrifugal terms appear representing the kinetic energy of the eliminated variables. While the Abelian gauge fixing via $\tilde{f}$ gives rise to a Coulomb-energy as in QED, implementation of the local gauge condition  (\ref{GGgc}) - $\delta f[\phi^{\alpha}]/\delta \alpha $ does not contain derivatives -   yields a local form of the centrifugal term. This locality makes the  Faddeev-Popov determinant factorize into contributions from every space-time point and the Gribov horizon consists of those configurations which contain matter fields with at least one zero. Here the analogy to quantum mechanics is complete with, for given $x$, the value of the adjoint scalar corresponding to the position of a particle moving in 3-space. In agreement with our general discussion,  the generating functional contains the non-trivial measure (\ref{rhomeas}) which is absent in  Hamiltonian which acts in the space of reduced, ``radial'' wave-functionals. 

\section{QCD in Coulomb-Gauge}

Subject of this section is QCD formulated in terms of transverse gauge fields as physical degrees of freedom. In particular we  will describe the implications of the compactness of these variables in Coulomb-gauge QCD. We follow here closely Gribov's pioneering work~\cite{GRIB78} in which the limitations in the range of the dynamical variables have been suggested as origin of confinement in Yang-Mills theories. The following discussion will be carried out within the canonical formulation with the expression (\ref{cgha}) as our starting point. An object of central interest is the interaction energy of two static color charges. In comparison with the corresponding expression in the Maxwell theory,  the external charges 
$$J^{0 a}(x) = g\rho^{a}({\bf x})=g\rho^{a}_{1}({\bf x})+g\rho^{a}_{2}({\bf x})$$
are coupled to the the charged gluons by the term ${\bf D}{\bf E}_{\mathrm{tr}}$. A corresponding coupling with similar consequences appears when a charged matter field is present in QED (cf.\ Eq.(\ref{quha})). On the other hand the modification of the Coulomb-propagator by the Faddeev-Popov operator is characteristic for a non-Abelian gauge theory.

\subsection{Asymptotic Freedom}

\begin{figure}
\noindent
\begin{center}
\begin{minipage}[b]{0.75\linewidth}
\epsfig{file=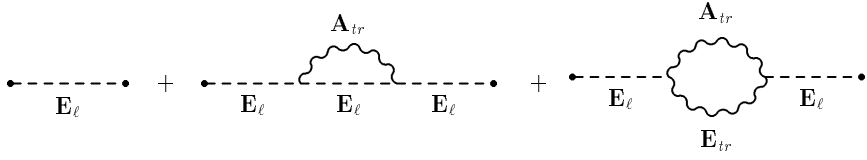, width=\linewidth}
\caption{Interaction energy of two static charges to one loop. Dashed lines indicate the instantaneous Coulomb-propagator, wavy lines transverse gluons.} \label{Fig1}
\end{minipage} \end{center}
\end{figure}

We begin our study of the dynamics  of the external charges by a perturbative calculation of the interaction energy~\cite{DREL}. This discussion not only will prepare the ground for the following non-perturbative considerations it will also illustrate how the characteristic antiscreening mechanism appears in a formulation in which only physical degrees of freedom and corresponding  positive norm states are present. For evaluation of the interaction energy in time independent perturbation theory, we introduce the total color charge density
\begin{equation}
  \label{rt}
\rho_{\mathrm{t}}^a({\bf x}) = \rho^a({\bf x}) - f^{abc}{\bf A}_{\mathrm{tr}}^b({\bf x})\,
{\bf E}^c_{\mathrm{tr}}({\bf x}).
\end{equation}
and write the Coulomb-energy contribution to the Hamiltonian (\ref{cgha}) in a QED like form
\begin{equation}
  \label{CE1}
\frac{1}{2}{\bf E}_{\mathrm{long}}^2 = -\frac{1}{2}\rho_{\mathrm{eff}}^a\,\frac{1}{\Delta}\,
\rho_{\mathrm{eff}}^a .
\end{equation} 
The expression for the effective charge density
\begin{equation}
  \label{CE2}
\frac{g}{\Delta}\,\rho_{\mathrm{eff}}= \frac{g}{{\bf D}\cdot\mbox{\boldmath$\nabla $}}\,\rho_{\mathrm{t}}
\end{equation} 
suggests to define the induced color charge density 
$$\rho_{\mathrm{eff}} = \rho_{\mathrm{t}} + \rho_{\mathrm{ind}}$$
with the perturbative expansion
\begin{eqnarray}
\rho_{\mathrm{ind}} &=& - ({\bf D}\cdot\mbox{\boldmath$\nabla $} - \Delta)\,
\frac{1}{\Delta + ({\bf D}\cdot\mbox{\boldmath$\nabla $}-\Delta)}\,\rho_{\mathrm{t}}
\nonumber \\
\label{ptex} & \approx & 
-gf^{abc}{\bf A}_{\mathrm{tr}}^b\cdot\mbox{\boldmath$\nabla $}\,
\frac{1}{\Delta}\rho_{\mathrm{t}}^c
+g^2f^{abc}{\bf A}_{\mathrm{tr}}^b\cdot\mbox{\boldmath$\nabla $}\,\frac{1}{\Delta}\,f^{cde}
{\bf A}_{\mathrm{tr}}^d\cdot\mbox{\boldmath$\nabla $}\,\frac{1}{\Delta}\rho_{\mathrm{t}}^e
+ \ldots
\end{eqnarray}
We now calculate to order $g^{4}$ the interaction energy. First order time independent perturbation theory yields contributions of arbitrarily high order in the coupling constant. In particular to order $g^{4}$ both the first and second order term in the expansion of $ \rho_{\mathrm{ind}}$ contribute and the following expression for the corresponding contribution to the interaction energy is obtained
\begin{eqnarray*}
\Delta E^{(1)} & = &-\frac{1}{2}\,g^2\int\!\rho_{\mathrm{eff}}^a\,
\frac{1}{\Delta}\,\rho_{\mathrm{eff}}^a\bigg|_{\mathrm{conn.}} \\
&=& - \frac{g^2}{2}\int\!\rho^a\,\frac{1}{\Delta}\,\rho^a  
- 3g^4f^{abc}f^{ade}\mat{0}
{\int(\mbox{\boldmath$\nabla $}\,\frac{1}{\Delta}\,\rho^e)\cdot{\bf A}_{\mathrm{tr}}^d\,\frac{1}{\Delta}
\,{\bf A}_{\mathrm{tr}}^b\,(\mbox{\boldmath$\nabla $}\,\frac{1}{\Delta}\,\rho^c)}
{0}_{\mathrm{conn.}}
\end{eqnarray*}
We have indicated in the formula that only connected interaction terms are retained. These $g^{2}$ and  $g^{4}$ contributions are given by the first two diagrams of Fig.~\ref{Fig1}. 
The standard expansion of the transverse gluon fields reads
\begin{equation}
  \label{aexp}
 {\bf A}_{\mathrm{tr}}^a({\bf x}) = \sum_{\lambda}\int\!d^3\!k\,\frac{1}{(2\pi)^{3/2}}\,\mbox{\boldmath$\epsilon $}({\bf k},\lambda)A^a({\bf k},\lambda)e^{i{\bf k}\cdot{\bf x}},
 \end{equation}
with the transverse polarization vectors $\mbox{\boldmath$\epsilon $}$ and the amplitudes $A^a$ which in turn are given in terms of  the creation and annihilation operators
\begin{equation}
  \label{aexp2}
A^a({\bf k},\lambda)=\frac{1}{\sqrt{2\omega_k}}\left( a^a({\bf k},\lambda)
+ a^a(-{\bf k},\lambda)^{\dagger}\right).
 \end{equation}
With $\omega_k=|{\bf k}|=k$, the gluon propagator in Coulomb gauge is obtained 
\begin{equation}
  \label{glpr}
\mat{0}{A_{\mathrm{tr}\,i}^a({\bf x})\,A_{\mathrm{tr}\,j}^b({\bf y})}{0}
= \int\frac{d^3\!k}{(2\pi)^3}\frac{1}{2\omega_{k}}\,
e^{i{\bf k}\cdot({\bf x}-{\bf y})}
\left( \delta_{ij} - \frac{k_ik_j}{ k^2} \right) \delta^{ab} \ ,
\end{equation}
and yields after Fourier-transform the following result 
\begin{eqnarray*}
\Delta E^{(1)} 
& = & 
g^2\int\!d^3\!k\,\rho_{1}^{a}({\bf k})\,\frac{1}{k^2}\,
\rho_{2}^{a}({\bf k})\left\{ 1 + \frac{3N_c g^{2}}{2\,(2\pi)^3k^2}\int\!d^3\!q\,
\frac{k^2 - ({\bf q}\cdot{\bf k})^2/q^2}{q\,|{\bf k}-{\bf q}|^2}\right\}\\
& = &
g^2\int \!d^3\!k\,\rho_{1}^{a}({\bf k})\,\frac{1}{k^2}\,
\rho_{2}^{a}({\bf k})
\left\{ 1 + \frac{12N_c g^{2}}{48\pi^2}\,\ln\frac{\Lambda^2}{k^2}\right\}.
\end{eqnarray*}
The logarithmic ultraviolet divergence has been regulated by cutting off the momentum integration at $\Lambda$. To order $g^{4}$ second order perturbation theory also contributes  
\beqs
\Delta E^{(2)} = 
\frac{1}{4}\,g^2\mat{0}{\int\!\rho_{\mathrm{t}}\,\frac{1}{\Delta}\,
\rho_{\mathrm{t}}\,\frac{Q}{E-H_0}\int\!\rho_{\mathrm{t}}\,
\frac{1}{\Delta}\,\rho_{\mathrm{t}}}{0}
 = - g^2\int\!d^3\!k\,\rho_{1}^{a}({\bf k})\,\frac{1}{k^2}\,
\rho_{2}^{a}({\bf k})
\,\frac{N_c g^{2}}{48\pi^2}\,\ln\frac{\Lambda^2}{k^2} \ .
\eeqs
The combined result for $\Delta E^{(1)}+ \Delta E^{(2)}$ leads to the well known expression for the running coupling constant
$$g^{2}(k)= \frac{g^{2}(\mu)}{ 1 + \frac{11N_c}{48\pi^2}\,g^2(\mu)\ln\frac{k^2}{\mu^{2}}} .$$  
Running of the coupling constant is a result of the two competing higher order contributions depicted in Fig.~\ref{Fig1}. The third term yields an increase in the coupling constant with increasing momentum. Arising in second order perturbation theory, the sign of this contribution is fixed irrespective of the dynamics; more generally, coupling to excited states always provides attraction; from this point of view it is clear that with  quarks taken into account this attractive contribution gets enhanced. In a gauge fixed formulation defined exclusively in terms of  physical degrees of freedom this universal attraction can be avoided only by processes such as given by the second diagram of Fig.~\ref{Fig1}. They involve the instantaneous interaction and cannot be interpreted as a coupling to physical states. 

\subsection{Gauge Fields Close to Gribov Horizons}

In using the standard form of the gluon propagator (\ref{glpr}) as basis of the perturbative calculation we have treated the transverse gauge fields as Gaussian variables. However, as has been pointed out by Gribov ~\cite{GRIB78}, in Coulomb-gauge and in Lorentz-gauge (as in many other gauges), the restriction to one gauge copy leads to a restriction in the domain of the dynamical variables, which invalidates application of standard perturbation theory. In various investigations subsequent to Gribov's work, the structure of the fundamental domain has been studied. In particular it can be seen by simple arguments~\cite{Zwan82} that the space of gauge field configurations within the Gribov region - the fundamental domain which includes  ${\bf A}_{\mathrm{tr}}=0$ - is bounded in all directions. We start with an arbitrary gauge field ${\bf A}_{\mathrm{tr}}$  in the Gribov domain, which necessarily satisfies
$$\int d^3 x \omega^{a\star}(x)\left(-\Delta\delta^{ab}-gf^{acb}\gamma {\bf A}_{\mathrm{tr}}^{c}(x)\mbox{\boldmath$\nabla$}\right)\omega(x)^{b}>0 . $$
In absence  of zero eigenvalues of the Faddeev-Popov operator  $-{\bf D}\mbox{\boldmath$\nabla$}$ inside the Gribov horizon and due to the positive definiteness of the Laplace operator the above integral cannot change sign when varying the constant $\gamma$ in the interval
$$ 0\le\gamma\le 1 .$$ 
On the other hand for sufficiently large $\gamma$ one always can achieve that the field dependent term in the Faddeev-Popov operator dominates. With the choice
$$\omega^{a}(x)= \rho(x)e^{i{\bf k}{\bf x}}v^{a}$$
with $\rho(x)\ge 0$ and  different from $0$ only in a region where none of the gauge field components $ A_{\mathrm{tr}\,i}^{a}$ changes sign. We furthermore assume that $|{\bf k}|$ is chosen large enough to make the variations of $\rho(x)$ negligible. In this way we  always can achieve  for  ${\bf k}$ of appropriate orientation  and for sufficiently large $\gamma$: 
\beqs
\int d^3 x \omega^{a\star}(x)\left(-\Delta\delta^{ab}-gf^{acb}{\bf A}_{\mathrm{tr}}^{c}(x)\mbox{\boldmath$\nabla$}\right)\omega(x)^{b}
\approx \gamma \,\mbox{Im}(v^{a\star}v^{b}) f^{abc}{\bf k}\int d^3 x  \rho(x)^{2}{\bf A}_{\mathrm{tr}}^{c}(x) <0. 
\eeqs
Thus for every ${\bf A}_{\mathrm{tr}}$ of the Gribov region the field configuration  $\gamma{\bf A}_{\mathrm{tr}}$ is beyond the Gribov horizon if $|\gamma|$ exceeds a certain finite value. 
Furthermore it can  be shown that the Gribov region is convex and  that the  Fourier components $ A^{a}({\bf k},\lambda)$  (Eq.(\ref{aexp2})) of a gauge field in the Gribov region  (finite volume $V$) are bounded by~\cite{AnZw89}  
\beqs
\frac{2\pi^{3}}{V^{2}} \sum_{{\bf k}, \lambda, a} \frac{\left| A^{a}({\bf k},\lambda)\right|^2 }{k^2} \le 180.
\eeqs
These basic insights into the geometry  of the Gribov region and further detailed~\cite{BAAL98} studies have not led to significant  progress concerning the dynamics since Gribov's  work. We therefore follow closely Gribov's original dynamical studies and  consider the expectation value of the inverse Faddeev-Popov operator
which has played such an important role in establishing asymptotic freedom in the Coulomb-gauge. We define
$$
G(k) = \langle0|\left({\frac{-1}{{\bf D}\cdot\mbox{\boldmath$\nabla $}}}\right)_{{\bf k}}|0\rangle =  \langle 0| \sum_{\lambda}\frac{\chi^{\star}_{\lambda}({\bf k},{\bf A}_{\mbox{tr}})\,\chi_{\lambda}({\bf k},{\bf A}_{\mathrm{tr}})}{\lambda[{\bf A}_{\mathrm{tr}}]} |0\rangle
$$
with  eigenvalues  $\lambda$ and eigenfunctions $\chi_{\lambda}$ of the Faddeev-Popov operator. This quantity is the ghost propagator which appears in the functional formulation if the Faddeev-Popov determinant is represented by a functional integral over Grassmann variables. After an appropriate ultraviolet regularization, $G(k)$ becomes infinite only if the Faddeev-Popov operator develops a zero eigenvalue. As discussed above this happens whenever the fundamental domain is left and more than one Gribov copy is taken into account. In a complete dynamical calculation this is prevented by the action of the potential ${\cal V}[A]$ to be included in the Hamiltonian and which becomes infinite beyond the Gribov horizon. We are not able to construct ${\cal V}[A]$; for a qualitative investigation of the effects of the limited domain of the dynamical variables one may require $G(k)$ to be finite and analyze the consequences of this requirement. This analysis can be carried out perturbatively by expanding the Faddeev-Popov operator around the Laplace operator in the same way as in Eq.(\ref{ptex}). Assuming  only color singlet operators to have non vanishing vacuum expectation values, the two leading terms in the expansion of $G(k)$ are (the system is defined in a region of volume~$V$)
\begin{eqnarray*}             
G(k)^{aa^{\prime}} \! \! \! \! \! &=& \! \! \! \! \langle 0|\left(\frac{-1}{\Delta}\right)_{{\bf k}}|0\rangle \delta^{a a^{\prime}} - g^2 f^{abc}f^{cb^{\prime}a^{\prime}}\langle 0|\left({\bf A}_{\mathrm{tr}}^{b}\cdot\mbox{\boldmath$\nabla $}
\frac{1}{\Delta}\,{\bf A}_{\mathrm{tr}}^{b^{\prime}}\cdot\mbox{\boldmath$\nabla $}\right)_{{\bf k}}|0\rangle + ... \\
&=& \! \! \! \! \delta^{a a^{\prime}} \left[\frac{V}{k^2}+g^{2}\langle 0|\left({\bf A}_{\mathrm{tr}}^{b}\cdot\mbox{\boldmath$\nabla $}\,
\frac{1}{\Delta}\,{\bf A}^{b}_{\mathrm{tr}}\cdot\mbox{\boldmath$\nabla $}\right)_{{\bf k}}|0\rangle +...\right] .
\end{eqnarray*}
Using the expansion (\ref{aexp},\ref{aexp2}) of the gluon field, we write by summing ladder graphs 
\begin{equation}
  \label{gper}
G(k)=\frac{V}{{\bf k}^2}\left(1 + \sigma(k,A)\right) +...
\approx\frac{V}{{\bf k}^2\,(1 - \sigma(k,A))}
\end{equation}
with 
\beqs
\sigma(k,A)= \frac{g^{2}}{V}\sum_{b,\lambda}\int\!\!\frac{d^3\!q}{(2\pi)^3}\,
\mat{0}{|A^b({\bf q},\lambda)|^{2}}{0}\,\frac{1}{({\bf k}-{\bf q})^2}
\left(1 - \frac{({\bf k}\cdot{\bf q})^2}{{\bf k}^2{\bf q}^2}\right).
\eeqs
We evaluate first $\sigma(k=0,A)$ in the perturbative vacuum;  ground state wave-functional and energy are given by
\beas
\psi_0[A] 
& = & 
\exp\left\{-\frac{1}{2 V}\sum_{{\bf k},\lambda,a} \,\omega_{k}
A^a({\bf k},\lambda)\,A^a(-{\bf k},\lambda)\right\}\nonumber \\
H_{0}\psi_0[A] 
& = & 
\frac{1}{2}\sum_{{\bf k},\lambda,a} \omega_{k}\,\psi_0[A], \quad \omega_{k}=|{\bf k}|
\eeas
with  
\beqs
 H_{0}= \frac{1}{2}\sum_{{\bf k},a,\lambda} \left(
-V\frac{\partial^2}
{\partial A^a({\bf k},\lambda)\,\partial A^a(-{\bf k},\lambda)}
+\frac{1}{V}\, {\bf k}^2\,A^a({\bf k},\lambda)\,A^a(-{\bf k},\lambda)\right).
\eeqs
Using the virial theorem, we read off the perturbative result for  $\sigma(k=0,A)$
$$
\sigma_0(0,A) 
= \frac{1}{3}\,\frac{g^2}{V}\sum_{{\bf k},\lambda,a}\frac{1}{k^3} .
$$ 
Thus in the perturbative vacuum $\sigma$ diverges logarithmically both in the infrared and ultraviolet. Independent of the approximations used, it is plausible that only after an ultraviolet regularization, a finite result can emerge. Scale invariance of the classical Lagrangian implies that Gribov horizons occur on all scales. With $\tilde{{\bf A}}_{\mathrm{tr}}(x)$ denoting a gauge field on the Gribov horizon, i.e.\ giving rise to a zero mode of the Faddeev-Popov operator $\chi_{0}(x)$,  the rescaled field $\lambda\tilde{{\bf A}}_{\mathrm{tr}}(\lambda x)$ also possesses a zero mode ($\chi_{0}(\lambda x)$) and thus  also belongs to the Gribov horizon. For making the above expression finite also an infrared cutoff would be required which indicates that in the true vacuum long wavelength gauge fields must be suppressed. 
Within the perturbative approach (Eq.(\ref{gper})) $\sigma(k,A)$ has to satisfy
$$\sigma(k,A) < 1 .$$
In order to characterize in a semiquantitative way this suppression of long-wavelength excitations we modify this condition. As we have seen, perturbative fluctuations favor large values of $\sigma$ and therefore one might expect the dominant contributions to arise actually from the region around the endpoint and we thus impose the constraint 
\begin{equation}
  \label{si1}
\sigma(0,A)= 1\, . 
\end{equation}  
As the final result will show,  the above constraint for $k\neq 0$ will be met provided the modified constraint (\ref{si1}) for $k=0$ is satisfied. We treat this condition by introducing a Lagrange multiplier $\bar{\kappa}$, i.e.\ the Hamiltonian is given by
\begin{equation}
  \label{csha}
 H_{\kappa}=H_{0} + \bar{\kappa}
\frac{4}{3}\,g^2\sum_{{\bf k}\lambda a} \frac{1}{|{\bf k}|^2}\,
A^a({\bf k},\lambda)\,A^a(-{\bf k},\lambda).\end{equation}
The substitution 
$$\displaystyle{{\bf k}^2\rightarrow{\bf k}^2 + \frac{\kappa^4}{{\bf k}^2}}\quad ,\quad \kappa^4 = \frac{4}{3}\,g^2\,\bar{\kappa}\,\frac{1}{V}$$
relates $H_{\kappa}$ and $H_{0}$; 
 the ground state energy of the constrained system is therefore
 \begin{equation}
   \label{EG}
E = \frac{1}{2}\sum_{{\bf k}\lambda a}\sqrt{{\bf k}^2 + \frac{\kappa^4}{{\bf k}^2}}\, . 
\end{equation}
The variational parameter $\bar{\kappa}$ is chosen such that the constraint is satisfied. Using once more the virial theorem we calculate
$$
\mat{0_\kappa}{A^a({\bf k},\lambda)\,A^{a}({\bf k},\lambda)}{0_\kappa}
=
\frac{k}{2\,\sqrt{k^4 + \kappa^4}}.
$$
The constraint (\ref{si1}) is satisfied provided $\kappa$ is given by
\begin{equation}
  \label{cst}
1 = \frac{2}{3}\, N_{c} g^2\int\!\frac{d^3\!k}{(2\pi)^3}
\frac{1}{k\sqrt{k^4 + \kappa^4}}=
\frac{2\,N_{c}\,g^2}{3\pi^2}\,\ln\frac{2\Lambda^2}{\kappa^2} \, ,\end{equation}
with $\Lambda$ regulating the ultraviolet divergence. This schematic calculation explicitly displays  suppression of the elementary excitations at long wavelengths as a result of the restriction in the range of integration of the gauge fields. The energies of the elementary excitations diverge with infinite wavelength (Eq.(\ref{EG})). 
Gaussian fluctuations of massless gluons would extend beyond the Gribov horizon and would lead to an overcounting of certain field configurations. This model calculation yields for  $G(k)$ the result (cf.~Eq.(\ref{gper})) 
$$\frac{{\bf k}^{2}}{V} G^{-1}(k) \approx 
1 - \sigma(k,A) =
 N_{c}g^2 \int\!\frac{d^3\!q}{(2\pi)^3}\,
\frac{k^2 - 2{\bf k}\cdot{\bf q}}{q\sqrt{q^4 + \kappa^4}}\,
\frac{1}{({\bf k}-{\bf q})^2}
\left( 1 - \frac{({\bf k}\cdot{\bf q})^2}{k^2q^2}\right)$$
with the long wave-length limit 
$$
1 - \sigma(k,A) \begin{array}[t]{c}
 \longrightarrow\\
 \raisebox{1ex} {\mbox{${\scriptstyle{ k \to 0}}$}}
 \end{array}
\frac{5\,N_{c} g^{2}}{3}\frac{k^2\ln \kappa^2/k^2}{\kappa^2}. 
$$
Thus at  infinite wave-length the ghost propagator diverges 
\begin{equation}
  \label{gk0}
 G(k)\begin{array}[t]{c}\longrightarrow\\
 \raisebox{1ex} {\mbox{${\scriptstyle{ k \to 0}}$}}
 \end{array} \frac{1}{k^{4}\ln \frac{\kappa^2}{k^2}}
\end{equation}
while  the gluon propagator (Eq.(\ref{glpr})) vanishes linearly with $k$ 
\beqs
\int d^3\!x e^{-i{\bf k}\cdot {\bf x}}\mat{0}{A_{\mathrm{tr}\,i}^a({\bf x})\,A_{\mathrm{tr}\,j}^b({\bf 0})}{0}
= \frac{1}{2\sqrt{{\bf k}^2 + \frac{\kappa^4}{{\bf k}^2}}}\,
\left( \delta_{ij} - \frac{k_ik_j}{ k^2} \right) \delta^{ab} . 
\eeqs
It is remarkable that a similar low momentum behavior has recently been obtained from a technically sophisticated calculation for  Lorentz gauge QCD within a Schwinger-Dyson approach~\cite{VSHA97,ALVS00} and this behavior also seems to be qualitatively confirmed by results from lattice QCD~\cite{BONN00,ALFF00}.

The ghost propagator is the essential quantity in the computation of the interaction energy of static charges (cf.~Eqs.(\ref{CE1},\ref{CE2})). If we neglect in a first step any polarization of the system when introducing a pair of static charges, this interaction energy is, within this model, given by
$$V({\bf x}_{1}-{\bf x}_{2}) \approx \frac{g^{2}}{(2\pi)^{3}}\int d^{3} k e^{i{\bf k}({\bf x}_{1}-{\bf x}_{2})} G^{2}(k) k^{2}$$
where we have used point  charges
$$\rho_{1,2}({\bf x}) = \delta({\bf x}-{\bf x}_{1,2}) .$$ 
The asymptotics of $V(r)$ at large separations of the static charges is obtained from  the long wave-length limit of $G(k)$ (Eq.(\ref{gk0})) 
$$V(r)\sim r^{3}.$$
The limited range in the integration over the gauge fields yields a confining interaction of static charges although the increase in energy with distance is too strong. The procedure of introducing static charges and neglecting any response of the system is too rough. The polarization of the medium should ultimately give rise to the formation of a flux tube and thereby lower the total energy of the system. Irrespective of the detailed mechanism, polarization  effects will lower the exponent in the interaction energy from this value $3$. A full calculation is, even within a harmonic approximation, technically rather involved. A qualitative idea about the polarization effects might be obtained by employing concepts from macroscopic electrostatics. As the perturbative calculation indicates, the essential polarization effect - the antiscreening leading to asymptotic freedom - is related to the appearance of the Faddeev-Popov operator in Eqs.(\ref{CE1},\ref{CE2}), only the much smaller screening contribution is due to the presence of the gluonic charge in $\rho_{t}$ (Eq.(\ref{rt})) which in the following will  therefore be neglected.  Eqs.(\ref{CE1},\ref{CE2}) suggest to introduce  the Maxwell displacement field ${\bf D}_{M}$ with $\rho_{\mathrm{t}}$ as its source while the source of the electric field is $\rho_{\mathrm{eff}}$, i.e.~${\bf E}$ is generated by by external and polarization charge. In this interpretation, the limited range of integration  gives rise to a dielectric constant $\epsilon$ which following the standard electrostatic definition 
$${\bf D}_{M} = \epsilon {\bf E}$$ 
is given by
$$ \epsilon(k) = \frac{G(k)}{k^{2}}.$$
Assuming such a linear relation, the energy required to separate two static charges in the polarizable medium of the Yang-Mills ground state is 
$$ V({\bf x}_{1}-{\bf x}_{2}) \approx \frac{1}{2} \int d^{3}x {\bf D}_{M}{\bf E}= \frac{g^{2}}{(2\pi)^{3}}\int d^{3} k e^{i{\bf k}({\bf x}_{1}-{\bf x}_{2})} G(k). $$
Polarization effects  reduce as expected the interaction energy and actually  lead to  a linear increase in the interaction energy  
$$V(r)\approx \sigma r$$
with the string constant determined by the parameter $\kappa$ (cf.~Eq.(\ref{cst}))  
$$\sigma = \frac{3\kappa^{2}}{80\pi N_{c}\ln \frac{\kappa r}{3}}\,\,.$$ 
The effect of the limited range of integration of the gauge fields due to the presence of a Gribov horizon leads naturally to dia-electric behavior of the QCD vacuum over large distances. Such behavior has been argued to be responsible for description of confinement in terms of bag-like models~\cite{TDLE}. Dia-electric behavior also implies, by covariance, that the magnetic permeability tends to infinity for small momenta and therefore points to an increasing  tendency of the QCD vacuum for spontaneous generation of magnetic fields at larger and larger scales. It thus appears that there is a deep connection between existence of Gribov horizons and magnetic instability of the QCD vacuum~\cite{SAV77}. It is however unlikely that, within the above approximations in treating the restrictions in the range of the gauge field variables, a qualitatively correct description of the magnetic field dominated QCD vacuum~\cite{NACH97} can emerge. We note that in the ground state of the Hamiltonian (\ref{csha}), the electric field energy dominates, we have
$$\langle 0|{\bf E}^{2}|0\rangle  = \frac{1}{4}\sum_{{\bf k}\lambda a}\sqrt{{\bf k}^2 + \frac{\kappa^4}{{\bf k}^2}}\, < \frac{1}{4}\sum_{{\bf k}\lambda a}\frac{k^{2}}{\sqrt{{\bf k}^2 + \frac{\kappa^4}{{\bf k}^2}}}\,= \,\langle 0|{\bf B}^{2}|0\rangle .$$
This model of QCD thus implies a departure from equipartition of electric and magnetic field energy. In contrast to the QCD vacuum, this modified vacuum is electric field dominated. This failure is most likely due to some inappropriateness in coping with the restrictions in the range of integration of the gauge field variables. In our discussion of the axial gauge a different  procedure is suggested in which compactness is actually the origin of the emergence of a magnetic vacuum.

\section{Axial Gauge QCD}

\subsection{Generating Functional and Hamiltonian}

The axial gauge choice singles out a certain direction in space-time. Therefore in this gauge
not only  manifest relativistic covariance is lost, as in any canonical formulation,  but in general also manifest rotational invariance. Nevertheless, for a variety of investigations and applications, the axial gauge representation of QCD is useful. For instance, the physics may single out a specific direction and thereby make the axial gauge choice particularly suited  such as in the evaluation of interaction energy of static charges. The axial gauge choice is the natural choice for  finite temperature QCD. In light-cone quantization the simplicity of the formalism is intimately related to the light-cone gauge, another type of axial gauge. For our discussion it is of particular importance that the gauge fixing procedure to  the axial gauge including the computation of the Faddeev-Popov determinant can be carried out explicitly and in closed form. In this way origin and  emergence of compact variables can be studied in detail. This in turn makes the physics consequences  transparent  and leads to various applications. In order to  properly define the axial gauge  one space-time direction has to be chosen compact. In this way certain ambiguities and associated infrared singularities  of QCD in the axial gauge are avoided. Here a spatial direction, the 3-direction, 
is assumed to be of finite extent $L$. Unlike the temporal gauge (time compact) this choice offers investigations of axial gauge QCD within canonical quantization without precluding applications to finite temperature QCD, as will be seen later. For this application it is important to assume the compact $x_{3}$ variables to be elements of a circle. We impose periodic boundary conditions for gauge fields and antiperiodic ones for fermion fields. In order to be specific we will restrict the discussion to SU$(2)$ QCD; generalization to SU$(N)$ is straightforward. The following gauge fixing procedure is very similar to the one employed in the above derivation of the unitary gauge representation of the Georgi-Glashow model.   

We write the axial gauge condition  as
\beq
\label{axgc}
f[A] = A_{3} - a_{3}\frac{\tau_{3}}{2}+\tilde{f}[A]
\eeq
i.e.\ in the transformation to axial gauge, the 3-component of the gauge field is replaced by a diagonal gauge field $a_{3}$. This field does not depend on $x_{3}$ and is compact 
\beq
\label{comvar}
  a_{3} = a_{3}(x_{\perp}),\quad 0\le a_{3}(x_{\perp})\le \frac{2\pi}{gL} \, .
\eeq
The notation $x=(x_{\perp},x_{3})$ has been used. As in the Georgi-Glashow model a subsidiary condition $\tilde{f}[A]$ has to be introduced in order to remove   invariance of $f[A]$ under Abelian, $x_{3}$-independent and in $x_{\perp}$ local, color-diagonal transformations. This residual redundancy can be eliminated  e.g.\ with the following choice  
\beq
\label{clb2}
\tilde{f}[A]= \sum_{i=1,2}\int _{0}^{L}dx_{3}\,\partial_{i}A_{i}^{3}\,\frac{\tau_{3}}{2} . 
\eeq
With the help of the auxiliary field $a_{3}(x_{\perp})$, the generating functional for QCD in axial gauge is written as 
\beqs
Z \left[J\right]= \int d[a_{3}]   
 \int d\left[A\right]\Delta_{f}\left[A\right]
\delta\left( A_{3} - a_{3}\frac{\tau_{3}}{2}\right) \delta( \tilde{f}[A])e^{i S \left[A\right]+i \int d^{4}x J^{\mu}  A_{\mu}}
\eeqs
The Faddeev-Popov determinant (cf.\ Eq. (\ref{FPD})) factorizes
$$ \Delta_{f}\left[A\right] = | \det D_{3} | \,\Delta_{\tilde{f}}\left[A\right]$$
and $\det D_{3}$ can be evaluated in closed form. The eigenvalue equation associated with $D_{3}$ for gauge fields satisfying the axial gauge condition (\ref{axgc}) reads
$$\left[\partial_{3}\delta^{ab}-g\epsilon^{ab3}a_{3}(x_{\perp})\right]\chi ^{b}_{n} = \lambda_{n}\chi^{a}_{n}$$
with  eigenfunctions $\chi(x)$ which are periodic in $x_{3}$. For $a=3$ and for the coupled system of the $a=1,2$ components, the eigenvalues are respectively 
\beqs
 \lambda_{n}^{3} = \frac{2i\pi n}{L},\quad\lambda_{n}^{\pm} = \frac{2i\pi n}{L}\pm ig a_{3}(x_{\perp}),\quad n=0,\pm 1,\pm 2,...\eeqs
and therefore the following value for the ratio of determinants (with zero modes of $\partial_{3}$ removed)
\beq
\label{det}
  \frac{\det D_{3}}{(\det \partial_{3})^{3}} = \prod_{x_{\perp}} g^{2}a_{3}^{2}(x_{\perp})\prod_{n=1}^{\infty}\left[1-\left(\frac{gLa_{3}(x_{\perp})}{2\pi n}\right)^{2}\right]^{2} 
= \prod_{x_{\perp}}\frac{1}{L^{2}}  \sin^{2} gLa_{3}(x_{\perp})/2   
\eeq
is obtained.  
The functional $\tilde{f}$ which fixes a residual U(1) gauge symmetry  is assumed to be linear; thus the corresponding contribution to the Faddeev-Popov determinant is field independent and we can write the generating functional as
\beq
\label{gfax1}
 Z \left[J\right]= \int D[a_{3}]   
 \int d\left[A_{\perp}^{\prime}\right]
\delta\left( \tilde{f}[A]\right) e^{i S \left[A_{\perp}^{\prime},a_{3}\right]+i \int d^{4}x J_{\perp}^{\mu}  A_{\perp \mu}} 
\eeq
where we have separated the measure 
$$d[A_{\perp}]=d[A^{\prime}_{\perp}] d[a_{3}]$$  
and we have absorbed the Faddeev-Popov determinant into the measure of the auxiliary field 
\beq
 D\left[a_{3}\right] = \prod_{x_{\perp}} \sin^{2}\left( gL
a_{3}(x_{\perp})/2 \right)\Theta \left(a_{3}(x_{\perp}) \right)\Theta \left( 2\pi/g L-a_{3}(x_{\perp}) \right)
 da_{3}\left(x_{\perp}\right).
\label{FE8}
\eeq
We conclude this purely formal derivation of the axial gauge representation of QCD with the construction of the Hamiltonian. The axial gauge condition gives rise to the following decomposition of the space of gauge fields (cf.\ Eq.(\ref{Mtr}))
\begin{eqnarray*}
  {\cal M}_{\perp}& =& \left\{{\bf E}_{\mbox{tr}}\right\}= \left\{{\bf E}\,\Big|\,(1-\delta_{a,3})E^{a}_{3}(x)\tau_{a}+\partial_{3}E^{3}_{3}(x)\tau_{3}= 0 \right\}\\{\cal M}_{\parallel} &=& \left\{{\bf E}\,\Big|\, E_{\mu}=0 ,\mu\neq 3, \,\,\int_{0}^{L} dx_{3}\,E_{3}(x) =0\right\}
\cup\left\{{\bf E}\,\Big|\,  \int_{0}^{L} dx_{3}\,(\partial_{1}E_{2}(x)-\partial_{2}E_{1}(x))=0 \right\}. 
\end{eqnarray*}
From Eqs.(\ref{axgc},\ref{clb2}) we obtain
$$ \frac{\delta f^{a}_{x}[A]}{\delta A_{i}^{b}(y)} = \delta_{i3}\,\delta^{ab}\,\delta(x-y) + \frac{1}{L}(1-\delta_{i3})\,\delta^{a3}\,\delta^{b3}\,\partial_{i}\,\delta(x_{\perp}-y_{\perp})$$ 
and using Eqs.(\ref{defg},\ref{ham}) the axial gauge Hamiltonian  can be derived
\bea
\label{hamaxg}
  H &=& \int d^{3}x \Big\{ \frac{1}{2} ( {\bf E}_{\perp}^2+{\bf B}[{\bf A}_{\perp}]^2 )  + {\bf J}_{\perp}{\bf A}_{\perp}
\nonumber \\
&& + \frac{1}{2} ({\bf D}\,({\bf E}_{\perp}+\mbox{\boldmath$\eta $})-J^{0})\frac{1}{(D_{3})^{2}}({\bf D}\,({\bf E}_{\perp}+\mbox{\boldmath$\eta $})-J^{0}) \Big\}.
\eea
The 2-dimensional longitudinal field appears when eliminating variables by implementing the subsidiary gauge condition $\tilde{f}$; it is given by 
$$\mbox{\boldmath$\eta$} (x_{\perp}) = \mbox{\boldmath$\nabla $}_{\perp}\frac{1}{L}\int_{0}^{L} dx_{3}\,({\bf D}_{\perp}\,{\bf E}_{\perp}-J^{0}\,)^{3}\frac{\tau^{3}}{2}\,  .$$

\subsection{Compact Variables}   

In this section we discuss the origin of the appearance  of compact variables in the elimination of redundant variables and indicate some consequences of the presence of such variables. Starting point of the above procedure is  the definition of gauge condition (\ref{axgc}). In order to justify this condition it remains to be shown that any arbitrary field configuration can be transformed by a gauge transformation into a configuration satisfying this condition. Given an arbitrary field $A(x)$, we pass to the axial gauge representation of QCD by applying the gauge fixing transformation  
\beq
  \Omega(x) = \Omega_{D}\left(x_{\perp}\right) (P^{\dagger}(x_{\perp}))^{x_{3}/L}
  \,
  P \exp\Bigl\{ig \int_{0}^{x_{3}} dz \,   A_{3}
    \left(x_{\perp},z\right)\Bigr\} \,
\label{cs2} 
\eeq
where 
\beq
\label{pol}
P(x_{\perp}) =   P \exp\Bigl\{ig \int_{0}^{L} dx_{3} \,   A_{3}(x) \Bigr\} 
\eeq
is the (untraced) Polyakov loop and the symbol $P$ denotes path ordering in the integration over $x_{3}$.
 The axial gauge is reached in 3 steps~\cite{LeNT94}. In the
presence of the third factor only, the gauge transformation would eliminate $A_{3}$
completely. In order to preserve the periodic boundary conditions of the gauge fields
the second term reintroduces zero mode fields. They in turn are diagonalized by
$\Omega_{D}$ which is defined by
\beq
\label{diag}
\Omega_{D}\left(x_{\perp}\right) P(x_{\perp})\Omega_{D}^{\dagger}\left(x_{\perp}\right) = e^{igL a_{3}(x_{\perp})\,\tau_{3}/2} \, .
\eeq
This gauge fixing procedure displays a peculiar  property of non-Abelian gauge theories, the appearance of group elements - here the Polyakov loops - in the process of elimination of Lie-algebra elements, the 3-component of the gauge field. The  only gauge invariant quantities which can be formed out of  a particular component of the gauge field are the eigenvalues of the Polyakov loop. In QED, it is the non-compact field 
$$ \mbox{QED:}\quad -\infty < a_{3}(x_{\perp}) < \infty $$ which is gauge invariant. A division of $a_{3}$ into a compact and an integer field would be  without consequences. The integers are the winding numbers which, for fixed $x_{\perp}$, are associated with the U(1) $\rightarrow$ U(1) mapping defined by the gauge field $A_{3}(x)$. These integers cannot be removed by gauge transformations. Physically, this reflects the fact that a non-compact field $a_{3}(x_{\perp})$ must be present, otherwise photons propagating in the $1-2$ plane would not exist in the Maxwell theory. In axial-gauge QCD, gluons polarized in the 3-direction  can propagate only as composite objects - built up from $A_{\perp}$ and the compact $a_{3}$.
These observations also imply a geometrical interpretation of the Faddeev-Popov determinant calculated in Eq.(\ref{det}). The resulting non-trivial measure in Eq.(\ref{FE8}) associated with the compact variables is the  Haar measure of  the group-elements $P(x_{\perp})$ and is therefore given  (for fixed $x_{\perp}$) by the  the volume element of the first polar angle on S$^{3}$. This interpretation also makes explicit the origin of the restriction in the range of integration (cf.\ Eq.(\ref{comvar})) of $a_{3}$.
As implied by  our general discussion, the Haar measure does not appear in the canonical formulation. The Hamiltonian acts on reduced wave-functionals which vanish at the  boundaries of the domain of definition
\beq
\label{abc}
\hat{\Psi}[A_{\perp}] = 0 \quad \mbox{if}\quad a_{3}(x_{\perp}) = 0, \frac{2\pi}{gL}.
\eeq
Full and reduced wave-functional  are related to each other by 
$$ \Psi[A_{\perp}] = \prod_{x_{\perp}}J^{-1}(a_{3}(x_{\perp}))\hat{\Psi}[A_{\perp}] , $$
where
$$J(a_{3}(x_{\perp})) = \sin gLa_{3}(x_{\perp}).$$
 The kinetic energy operator associated with the compact variables acting on the reduced wave-functional  
$$ \hat{K}_{a_{3}}\hat{\Psi}[A_{\perp}] =- \sum_{x_{\perp}}\frac{\partial^{2}}{\partial a_{3}(x_{\perp})^{2}}\hat{\Psi}[A_{\perp}]$$
is transformed into
 $$ K_{a_{3}}\Psi[A_{\perp}] =- \sum_{x_{\perp}}\frac{1}{J(a_{3}(x_{\perp}))}\frac{\partial}{\partial a_{3}(x_{\perp})}J(a_{3}(x_{\perp}))\frac{\partial}{\partial a_{3}(x_{\perp})}\Psi[A_{\perp}] \, ,$$
as is appropriate for the first polar angle on S$^{3}$. Using the full instead of the reduced wave-functional, the boundary condition (\ref{abc}) follows from the above form of the kinetic energy operator $K_{a_{3}}$. In agreement with our general remarks,  the last term of the axial gauge Hamiltonian in (\ref{hamaxg})  is to be interpreted as the centrifugal energy which appears in the separation of the angular variables of S$^{3}$. Indeed as can be seen from the explicit form of the eigenvalues, $D_{3}^{-2}$ develops a second order pole when $a_{3}(x_{\perp})$  approaches the boundaries  $0$ or $  \frac{2\pi}{gL} $ of the domain of definition.\\
 Here we have used functional methods to derive the gauge fixed formulation of QCD. It is reassuring that the results are the same as obtained in canonical quantization~\cite{LeNT94}. Canonically one quantizes QCD in the Weyl gauge $A_{0}=0$ and subsequently  implements the Gau\ss\ law in order to reach  the gauge fixed formulation; this procedure actually leads to the Hamiltonian corresponding to the full wave functional $\hat{\Psi}[A_{\perp}]$. In turn, using this canonical formulation as starting point, the generating functional  can be derived and  issues concerning the compact variables as discussed above have been studied from a different perspective~\cite{JAKS97} in 1+1 dimensional QCD.

\subsection{Polyakov-Loops and Center Symmetry}

Our discussion has so far concentrated on the origin of compact variables and their impact on the structure of the resulting gauge fixed formulation. The compact variables appearing in axial gauge, the Polyakov loop variables (\ref{pol}) have a significant dynamical role. Their eigenvalues serve as order parameters for the confined and deconfined phases of pure Yang-Mills theories~\cite{McLerran81,Svetitsky86,Lenz98}. Expectation value and correlation functions of these variables are related to the self energy of a single static quark  and the interaction energy of two static quarks respectively and therefore distinguish for instance the gluon plasma from the  confining phase. The Polyakov loops acquire this dynamical significance for symmetry reasons. As will be shown now, the gauge fixing procedure discussed above is actually not complete. A discrete residual gauge symmetry, the center symmetry, is still present and the Polyakov loop distinguishes the realization of this symmetry. 
Under gauge transformations $\Omega(x)$, $P\left(x_{\perp}\right)$ transforms as
\beqs P(x_{\perp}) \rightarrow
\Omega\left(x_{\perp},L\right)P(x_{\perp})
 \Omega^{\dagger}\left(x_{\perp},0\right)\ .
\eeqs 
The coordinates  $x=(x_{\perp},0)$ and
$x=(x_{\perp},L)$ describe identical points and we require the
periodicity properties imposed on the field strengths not to change
under gauge transformations. This is achieved if
$\Omega$  satisfies
\beqs
 \Omega\left(x_{\perp},L\right) = c_{\Omega}\cdot \Omega\left(x_{\perp},0\right)
\eeqs with $c_{\Omega}$ being an element of the center of the
group. Thus gauge transformations can be classified according to the
value of $c_{\Omega}$ ($\pm 1$ in SU(2)). Therefore under gauge
transformations
\beqs
\mbox{tr}(P(x_{\perp})) \rightarrow
\mbox{tr}(c_{\Omega} P(x_{\perp}))
\stackrel{\mathrm{SU(2)}}{=} \pm \mbox{tr}(P(x_{\perp})) .
\eeqs 
 A simple example of an SU(2) gauge transformation $\omega_{-}$
with $c=-1$ is 
\beq
\omega_{-} = e^{i\pi \mbox{\boldmath$\hat{\scriptstyle\psi}
$}\mbox{\boldmath$\scriptstyle\tau $}x_{3}/L}\quad, \quad c_{\omega_{-}}=-1 .
\label{cs1a}
\eeq
with the arbitrary unit vector $\hat{\psi}$. Its effect on an arbitrary gauge field is
\beqs
  A_{\mu}^{\,[\omega_{-}]} = e^{i\pi \mbox{\boldmath$\hat{\scriptstyle\psi}
$}\mbox{\boldmath$\scriptstyle\tau $} x_{3}/L}A_{\mu} 
e^{-i\pi \mbox{\boldmath$\hat{\scriptstyle\psi}
$}\mbox{\boldmath$\scriptstyle\tau $} x_{3}/L}-\frac{\pi}{gL}\mbox{\boldmath$\hat{\psi}
$}\mbox{\boldmath$\tau $}\delta_{\mu 3}. 
\eeqs
This representative $\omega_{-}$ can
be used to generate any other gauge transformation changing the sign of $\mbox{tr}(P)$ by multiplication with a
strictly periodic ($c=1$) but otherwise arbitrary gauge transformation. 
 The decomposition of SU$(2)$ gauge
transformations into two classes according to $c=\pm 1$ implies a
decomposition of each gauge orbit $\cal O$ into sub-orbits
 $\cal O_{\pm}$ which are characterized by the sign of the Polyakov
loop at some fixed $x_{\perp}^{0}$
\beqs
 A(x)\, \in \, {\cal O}_{\pm} \quad ,\quad \mbox{if}\quad \pm
\mbox{tr}(P (x_{\perp}^{0}))\ge 0 .
 \eeqs 
Thus strictly speaking, the trace of the Polyakov loop
is not a gauge invariant quantity. Only $|\mbox{tr}(P(x_{\perp})|$
is invariant under all  gauge transformations. Furthermore the
spontaneous breakdown of the center symmetry in Yang-Mills theory as it
supposedly happens in the transition from the confined to the quark-gluon plasma phase constitutes a
breakdown of the underlying gauge symmetry. This implies that the wave
functional describing such a state is different for gauge field
configurations which belong to $\cal O_{+}$ and $\cal O_{-}$
respectively, and  which therefore are  connected by gauge
transformations such as $\omega_{-}$ in Eq. (\ref{cs1a}). These symmetry
considerations apply equally well when adjoint matter is coupled to the  gauge fields. They have been applied to 1+1 dimensional QCD with adjoint fermions~\cite{LeST95} and have shown to be useful in the characterization of the various phases of the Georgi-Glashow model~\cite{LeST95,LNRT00}. On the other hand, with matter in the fundamental representation  present, the system is not anymore invariant under center symmetry transformations, which change the boundary condition of fields carrying
fundamental charges.\\
In general we have to expect the center symmetry transformations to be present after gauge fixing. More precisely, whenever gauge fixing is carried out exactly and with strictly periodic gauge fixing transformations ($\Omega, c_{\Omega}=1$)
 the resulting formalism must contain transformations which change the Polyakov loop  
\beqs
\mbox{tr}(P(x_{\perp})) \rightarrow  -\mbox{tr}(P(x_{\perp})) .
\eeqs  
A discrete residual gauge symmetry is left and  each gauge
orbit is represented by two gauge field configurations.
The gauge fixing transformation to the axial gauge (\ref{cs2}) is periodic and therefore residual symmetry transformations must be present. It is straightforward to construct these  transformations, the  center reflections. We use the following definition 
\beqs Z =  i   e^{i\pi \scriptstyle\tau_{1}/2} e^{i\pi
\tau_{3} x^{3}/L} .
\eeqs
These transformations
are reflections and change the sign of the Polyakov loop
\beqs
  Z^{2}=1,\quad c_{Z}=-1\, .
\eeqs 
They do not change the gauge condition (\ref{axgc}) and are therefore symmetry transformations of the gauge fixed generating functional (\ref{gfax1}) and Hamiltonian (\ref{hamaxg}).

The effect of Z  on  arbitrary gauge fields can be simplified by representing the fields in a spherical color basis 
\beq
\Phi_{\mu}(x)= \frac{1}{\sqrt{2}}(A_{\mu}^{1}(x)+i A_{\mu}^{2}(x))e^{-i\pi x^{3}/L}
\label{gf12a} 
\eeq
and by shifting the Polyakov loop variables
\beq
\label{shift}
a_{3}(x_{\perp})-\frac{\pi}{gL}  \rightarrow  a_{3}(x_{\perp}).
\eeq
As suggested by this definition we will refer to $\Phi_{\mu}$ as charged and to $A^{3}_{\mu}$ as neutral components  of the gauge field.
 Under  center reflections, these fields transform as
\beq
\label{gf12b}
 Z: \quad a_{3} \rightarrow -a_{3}\quad ,\quad A_{\mu}^{3} \rightarrow
-A_{\mu}^{3}\quad ,\quad \Phi_{\mu} \rightarrow  \Phi_{\mu}^{\dagger} \quad ,\quad
(\mu \neq 3 ).
\eeq 
and therefore  the trace of the Polyakov loop changes sign  
\beq
\label{trpo}
\mbox{tr}\, P(x_{\perp}) = \sin gLa_{3}(x_{\perp}),\quad  Z: \quad \mbox{tr}\, P(x_{\perp}) \rightarrow -\mbox{tr}\, P(x_{\perp}).
\eeq
The center symmetry transformation $Z$ acts as (Abelian) charge
conjugation with the ``photons'' described by the neutral fields $A_{\mu}^{3}(x),
a_{3}(x_{\perp}) $. As will be seen
shortly these field redefinitions will also simplify the description of the dynamics. The phase
change in Eq.(\ref{gf12a}) makes  the charged fields antiperiodic
\beq
\Phi_{\mu}(x_{\perp}, x^{3}=L) = - \Phi_{\mu}(x_{\perp}, x^{3}=0) . 
\label{cs3c} 
\eeq
If the center symmetry is realized  $ gL a_{3}({\bf x_{\perp}})$ has to be
distributed symmetrically around the origin.
With the Polyakov loop defined with respect to a spatial compact direction, the center reflections are standard symmetry transformations of the canonical theory. In particular, we can associate with Z an operator which commutes with the gauge fixed Hamiltonian of Eq.(\ref{hamaxg})

$$[H,Z] = 0 \,.$$
As a consequence, if the center symmetry is realized, the stationary states can be classified as Z-even or Z-odd states.

We conclude this section with some general remarks concerning the relation between covariant field theories at finite extension and finite temperature. For the formulation of the center symmetry and
definition of the Polyakov loops we have considered QCD  in a geometry
where the system is of finite extent $L$ in a spatial  direction $x_{3}$ in order to be able to interpret the center symmetry in the canonical formalism. In finite temperature QCD, such a standard interpretation of the center reflections as symmetry transformations is not
possible and  conceptual difficulties may arise~\cite{SMIL94}
concerning for instance the existence of domain walls. However these difficulties can be circumvented. By covariance,  relativistic field theories at finite (spatial) extension and finite temperature are related to each other.  
By rotational invariance in the Euclidean the value of the partition function of a system
with finite extension $L$ in 3 direction and $\beta$  in 0 direction is invariant under
the exchange of  these two extensions
\beqs Z\left(\beta,L\right)=Z\left(L,\beta\right) \ ,
\eeqs provided standard boundary conditions in both time and 3 coordinate are imposed on the fields. As a consequence energy
density and pressure are related by
\beq
\epsilon\left(\beta,L\right)=-p\left(L,\beta\right) \ .
\label{FE2}
\eeq
For a system of non-interacting particles this relation connects energy
density or pressure of the Stefan-Boltzmann law with the corresponding quantities
measured in the Casimir effect.
In QCD, covariance  implies by Eq.(\ref{FE2}) that at zero temperature a
confinement-deconfinement transition occurs when compressing the QCD vacuum (i.e.\ 
decreasing $ L$). From lattice gauge calculations~\cite{Kanayo} it can be inferred that
this transition occurs at a critical extension $L^{c} \approx 0.8$ fm in the absence of
quarks and at $L^{c} \approx 1.3$ fm when quarks are included. For extensions smaller
than
$L^c$, the energy density and pressure reach values which are typically 80 \% 
of the corresponding  ``Casimir'' energy and pressure. When compressing the system beyond
the typical length scales of strong interaction physics, correlation functions at
transverse momenta or energies $|p| \ll 1/L $ are dominated by the zero ``Matsubara
wave-numbers" in 3-direction and, as confirmed by lattice QCD calculations~\cite{Reisz},
are  given by the dimensionally reduced QCD$_{2+1}$.  

\subsection{Gauge Ambiguities and Monopoles}
Apart from the discrete center symmetry transformation described by the charge
conjugation $Z$, all other symmetries related to the gauge invariance have
been used to eliminate $A_{3}$. In such a case of a global, non-perturbative gauge fixing we have to expect, as argued above, exceptional field configurations
to emerge. In transforming to the axial gauge, diagonalization of the Polyakov loops ($\Omega_{D}$ in
Eq.(\ref{cs2}))  is the crucial step of the gauge fixing procedure, in which
such singular gauge
field configurations appear. 
The diagonalization (\ref{diag}) can also be interpreted as a choice of
coordinates in color space. With Eq.~(\ref{diag}), the color 3-direction is
chosen to be that of the Polyakov loop at given $x_{\perp}$. For studying the emerging singular fields it is convenient to introduce
polar ($\theta $) and azimuthal ($\varphi $) angles in color space,
\beqs
 P(x_{\perp})= p_{0}(x_{\perp})\, {1\mkern-4mu\rm l} +i{\bf p}(x_{\perp})\mbox{\boldmath$\tau $}  
\eeqs
with
\beas
p_{0}(x_{\perp})&=&\cos(\frac{1}{2} gL a_{3}\left(x_{\perp}\right))
\nonumber\\
{\bf p}(x_{\perp})\mbox{\boldmath$\tau $}&= & \sin(\frac{1}{2} gL a_{3}\left(x_{\perp}\right)) \left(
    \begin {array}{cc} 
      \cos(\theta \left(x_{\perp}\right)) &
      e^{-i\varphi  \left(x_{\perp}\right)}\sin(\theta \left(x_{\perp}\right)) \\
      \noalign{\medskip}
      e^{i\varphi  \left(x_{\perp}\right)} \sin(\theta \left(x_{\perp}\right)) &
      - \cos(\theta \left(x_{\perp}\right))
    \end {array} \right) .
\eeas
  With this choice of coordinates, the
matrix $\Omega_{\olabbr{D}}\left(x_{\perp}\right)$ can be represented as
\beqs 
  \Omega_{\olabbr{D}}\left(x_{\perp}\right) =
  \left (\begin {array}{cc}
      e^{i\varphi  \left(x_{\perp}\right)} \cos(\theta \left(x_{\perp}\right)/2) &
      \sin(\theta \left(x_{\perp}\right)/2) \\\noalign{\medskip}
      - \sin(\theta \left(x_{\perp}\right)/2) &
      e^{-i\varphi  \left(x_{\perp}\right)} \cos(\theta \left(x_{\perp}\right)/2)
    \end {array} \right) .
\eeqs
In general, diagonalization or equivalently the choice of coordinates is not
everywhere well defined and consequently (coordinate-)singularities occur in
the associated transformations of the gauge fields.  Starting from an
everywhere-regular gauge field $A_{i}$ the transformed field
\beq 
  A_{\mu}^{\prime} \left(x\right) = \Omega_{\olabbr{D}}\left(x_{\perp}\right) 
  A_{\mu} \left(x\right)\Omega_{\olabbr{D}}^{\dagger}\left(x_{\perp}\right)
  + s_{\mu}\left(x_{\perp}\right)  
\label{am5} 
\eeq
with 
\beq 
  s_{\mu}\left(x_{\perp}\right) = \Omega_{\olabbr{D}}\left(x_{\perp}\right) 
  \frac{1}{ig}\partial _{\mu}\Omega_{\olabbr{D}}^{\dagger}\left(x_{\perp}\right) 
\label{am5a} 
\eeq
is in general singular with $\Omega_{\olabbr{D}}$.  While the homogeneous term
can at most be discontinuous the inhomogeneous term diverges. We note that $ s_{\mu}\left(x_{\perp}\right)$ is determined exclusively by the Polyakov-loop variables $p_{0},{\bf p}$ and  furthermore it can be shown that the parameters characterizing the singularities are given in terms of the gauge invariant eigenvalues of the Polyakov loops $p_{0}$. We represent  the  inhomogeneous term $ s_{\mu}$ in a spherical color basis (cf.\ Eq.(\ref{gf12a})) 
\beqs 
  s_{\mu}\left(x_{\perp}\right)  =  a_{\mu}^{\olabbr{s}}\left(x_{\perp}\right)\tau_{3}
  +\frac{1}{\sqrt{2}}\, (\phi_{\mu}^{\olabbr{s}}\left(x_{\perp}\right) (\tau_{1}-i\tau_{2})+\mbox{h.c.}) ,
\eeqs
where
\bea
   \phi_{\mu}^{\olabbr{s}}(x_{\perp}) & = & \! \frac{1}{2g} e^{-i\varphi  }
   \left(\sin\theta \, \partial_{\mu}\varphi  
     - i\partial_{\mu}\theta  \right) =
   \frac{1}{2g\sin{gL a_{3}}} e^{-i\varphi} \,
   ( \hat{\mbox{\boldmath$\varphi $}}-i \hat{\mbox{\boldmath$\theta$}})
   \cdot \partial_{\mu}\,{\bf p}\left(x_{\perp}\right)
\nonumber\\
\label{gf14}
   a_{\mu}^{\olabbr{s}}(x_{\perp}) & = & \! -\frac{1}{2g}\left(1+\cos\theta \right)
   \partial_{\mu}\varphi   =
   - \frac{1+\cos\theta }{2g\sin{\theta }}
   \frac{1}{\sin{gL a_{3}}} \,
   \hat{\mbox{\boldmath$\varphi $}} \cdot \partial_{\mu}\,{\bf p}\left(x_{\perp}\right) ,
\eea
with the standard choice of unit vectors (in color space)
\beqs
   \hat{\mbox{\boldmath$\varphi $}}=\left(
    \begin{array}{c}
      -\sin{\varphi  }\\
      \cos{\varphi  }\\
      0
    \end{array}\right)
  \qquad \mbox{and} \qquad
   \hat{\mbox{\boldmath$\theta $}} = \left(
    \begin{array}{c}
      \cos{\theta }\cos{\varphi  }\\
      \cos{\theta }\sin{\varphi  }\\
      -\sin{\theta }
    \end{array}
  \right).
\eeqs
These expressions display the nature of the singular fields and describe the conditions under which singularities occur in the process of gauge fixing. We assume  the  gauge field $A_{\mu}(x)$ and in particular the corresponding Polyakov loop $P =(p_{0},{\bf p})$  to be smooth before diagonalization. Singularities (poles) occur at points
$x_{\perp}^{\olabbr{N},\olabbr{S}}$,
where the Polyakov loop passes through the center of the group and does not
define a direction in color space,
\beq
z\equiv P(x_{\perp}^{\olabbr{N},\olabbr{S}})= \pm {1\mkern-4mu\rm l} .
\label{za2}
\eeq
This happens if 
\beqs
 gL a_{3}( x_{\perp}^{\olabbr{N}}) = 0\quad \mbox{or} \quad gL a_{3}( x_{\perp}^{\olabbr{S}}) = 2\pi 
\eeqs
i.e.\ if the Polyakov loop variable reaches the border of the fundamental domain (Eq.(\ref{comvar})). 
This requirement determines a point on the group manifold $\mbox{S}^{3}$ and
thus, for generic cases, fixes (locally) uniquely the position
$x_{\perp}^{\olabbr{N},\olabbr{S}}$.  In 4-space the transformed gauge fields are thus
singular on straight lines parallel to the 3-axis. As a consequence of the particular gauge condition involving the diagonalization of the $x_{3}$ independent Polyakov loop, they are also independent of $x_{3}$ and have a vanishing $\mu=3$ component.   The singularities can be classified according to the value of the Polyakov loop and we shall refer to them as north and south pole singularities according to
the respective positions of the Polyakov loop on the group manifold SU(2)
($\cong{\rm S}^3$).  Thus we can assign a ``north-south'' quantum number or
charge $z$ to this singularity (i.e.\ the range of $z$ is the center of the
group).   In addition to
poles, the field $a_{\mu}^{\olabbr{s}}$ also exhibits (static) string like
singularities along the line $\theta(x_{\perp}) =0$ representing a surface in 4-space. The
charged gluon fields too, have poles at $x_{\perp}^{\olabbr{N},\olabbr{S}}$ and
discontinuities along the strings $\theta(x_{\perp})=0$.  Although the points
$x_{\perp}^{\olabbr{N},\olabbr{S}}$ are characterized by Eq.(\ref{za2}) in a gauge
invariant way (by the degeneracy of two gauge invariant eigenvalues), in
general those points have no particular significance in other gauges.
To illustrate this general discussion we consider a particularly simple singular field which arises if we identify color orientation and spatial orientations. For this we assume the $x_{\perp}$ space to be Euclidean (imaginary time) and we use a vector notation (e.g.\ ${\bf a}=(a_{0},a_{1},a_{2})$). Identification of color and spatial orientation 
\beqs 
  \varphi \left(x_{\perp}\right) =  \varphi_{0}\equiv   \arctan \frac{y}{x}
  \quad ,\quad
  \theta \left(x_{\perp}\right) = \theta_{0}\equiv\arccos \frac{t}{\sqrt{x^2+y^2+t^2}}\, .
\eeqs
generates a singular field $a_{i}^{\olabbr{s}}(x_{\perp})$ in Eq.(\ref{gf14})
whose  neutral component is the vector potential of a Dirac monopole~\cite{DIRAC31} of
charge $2\pi/g$,
\beq
  {\bf a} \left(x_{\perp}\right)  
  = - \frac{1}{2g}\frac{1+ \cos\theta_{0}}{r \sin\theta_{0}}\,
  \hat{\mbox{\boldmath$\varphi $}}_{0} 
\label{am18}
\eeq
with Abelian (neutral) magnetic field
\beqs
  {\bf b} = {\rm curl}\, {\bf a} 
  =\frac{1}{2g} \frac{{\bf x}_{\perp}}{x_{\perp}^3} .
\eeqs
Associated with the singular neutral component is a singular charged component
 which is given by
\beq
 \mbox{\boldmath$\phi $} \left(x_{\perp}\right) = -\frac{1}{2gr}
  \left(\hat{\mbox{\boldmath$\varphi $}}_{0}+i\hat{\mbox{\boldmath$\theta$}}_{0}\right)
  e^{- i \varphi_{0}} .
\label{za3}
\eeq

 This example exhibits a general property of the singular field configurations generated by the diagonalization. The strength of the singularity of the neutral component is determined by the winding of the color orientation when the orientation in $x_{\perp}$ space is appropriately varied. The magnetic charge $m$ is quantized and given by
$$ m = \frac{2\pi n}{g},\quad n=\pm1,\pm2,... \ .$$ 
On the other hand, the structure of the singular charged component is, in general,  not determined  exclusively by topological properties. However the singularities of neutral and charged components are intimately related. The Abelian field strength corresponding to the neutral singular fields, a central quantity of the so called Abelian projection
\beqs
f_{\mu \nu} = \partial_{\mu}a_{\nu}^{\olabbr{s}}-\partial_{\nu}a_{\mu}^{\olabbr{s}} \ , \quad \mu,\nu\neq 3
\eeqs
 is singular at the
position of the monopoles. On the other hand, the complete non-Abelian field
strength built from the inhomogeneous term $s_{i}$ of Eq.(\ref{am5a}) actually
vanishes,
\beqs
  F_{\mu \nu}\left[s\right] =
  \partial_{\mu}s_{\nu}-\partial_{\nu}s_{\mu}+ig\left[s_{\mu},s_{\nu}\right] = 0 ,
\eeqs
i.e.\ the singular Abelian field strength is exactly canceled by the
non-Abelian contribution to $F$ generated by the singularities in the
charged gluon fields. Thus ``Abelian'' monopoles have finite or possibly even vanishing field strength. Singularities in the gauge fields necessarily cancel in gauge invariant quantities; they have been produced as coordinate singularities  since we
insisted on fixing the gauge globally. Such cancellations can be achieved only if the connection between neutral and charged singular components
of the gauge fields contained in the above expressions (\ref{am5},\ref{gf14}) is not disturbed.
Cancellation of the singularities in
the Abelian field strength by the non-Abelian commutator must happen quite
generally,  the gauge fixing procedure  cannot
affect gauge invariant quantities even at the positions of the monopoles and
along the strings. 
 The above considerations also make clear that, in general, there are no bounds on the action associated with singular fields. The simple example of the gauge field
$$A_{\mu}(x) = \frac{\pi}{gL}\,\tau_{3}\,\delta_{\mu 3}$$  
illustrates this point. The corresponding action is zero and the whole space will be filled with monopoles after transforming to axial gauge. 
On the other hand, fluctuations around singular field configurations will disturb the delicate balance between Abelian and non-Abelian contribution to the field strength and, in general, an infinite action will result. Fluctuations have to satisfy specific requirements to yield finite action. 
The charged components of the fluctuations $\delta \phi$ have to satisfy the condition
\beq
\label{str}
 \delta\phi(x)\, e^{2i\varphi(x_{\perp})} \quad \mbox{continuous along the strings}
\eeq
and both neutral and charged fluctuations have to vanish at the position of the pole.

Unlike the points $x_{\perp}^{\olabbr{N},\olabbr{S}}$ which are determined in a gauge
invariant way (cf.\ Eq.(\ref{za2})) the location of the strings is to a large
extent  arbitrary and depends on the details of the subsidiary gauge condition $\tilde{f}[A]$ of Eq.(\ref{axgc}). An Abelian rotation with an
$x_{\perp}$-dependent gauge function only affects the $\tilde{f}[A]$ part of the gauge condition.   This subsidiary gauge condition can be used to simplify the description of the strings. 

For the following discussion and for further applications it is convenient to make explicit in the generating functional the contributions from singular field configurations. This is not   necessary if a complete evaluation of the path integral is attempted. For approximative evaluation however a decomposition into singular fields and fluctuations is useful. To this end, we  classify the gauge fields according to the number
${\bf n}=(n_{N},n_{S})$ of north ($n_{N}$) and south ($n_{S}$) pole
singularities i.e.\  according to the number of times the Polyakov loop passes through the ``north'' and ``south'' center respectively and denote the corresponding fields by $a_{3}^{{\bf n}}$ . On the basis of this classification the generating functional can be decomposed as
\beq
Z[J]= \sum_{{\bf n}}Z_{{\bf n}}[J] =
\sum_{{\bf n}} \int D[a_{3}^{{\bf n}}]   
  d\left[A_{\perp}^{\prime}\right]
\delta\left( \tilde{f}[A]\right) e^{i S \left[A_{\perp}^{\prime},a_{3}^{{\bf n}}-\pi/gL\right]+i \int d^{4}x J^{\mu}  A_{\perp \mu}} 
\label{pm0a} 
\eeq
As in Eq.(\ref{gfax1}) the integration variables, the unconstrained degrees of freedom, are the 3
components of the gauge field $A_{\perp}^{\prime} =\left\{A_{\mu}(x), \mu \neq 3\right\}$ and the eigenvalues of the Polyakov loops.    For ${\bf n}\neq 0$ singularities show up in the gauge field components $A_{\perp}^{\prime}$ which in turn are determined by the $x_{\perp}$ dependence of $a_{3}^{\bf n}$. One therefore can split $A_{\perp}^{\prime}$ into  singular  and fluctuating fields
\beq
\label{split}
A_{\mu}^\prime(x) = s_{\mu}(x_{\perp}) +\delta A_{\mu}^\prime(x).    
\eeq
For given $a_{3}(x_{\perp})$, the singular fields
\beqs
  s = s \left[a_{3}^{{\bf n}}\right] .
\eeqs
 can be constructed as generalization of the Dirac monopole solutions (Eqs.(\ref{am18},\ref{za3})). For completeness we also write down the  measure (cf.\ Eq.(\ref{FE8})) of the Polyakov loop variables, which after the field redefinition (\ref{shift}) reads
\beq 
D\left[a_{3}\right] = \prod_{x_{\perp}} \cos^{2}\left( gL
a_{3}(x_{\perp})/2 \right)\Theta \left( \frac{\pi^{2}}{g^2 L^2} -a_{3}^2 (x_{\perp}) \right)
 da_{3}\left(x_{\perp}\right).
\label{FE8p}
\eeq

\subsection{Phenomenology of Polyakov-Loops}

Up to this point our discussion has focused on the development of the formalism for gauge fixed theories and a description of the properties of the physical variables reached after gauge fixing. Within axial gauge QCD, it has been possible to carry out a rather complete analysis of the properties of the physical, unconstraint variables. We have demonstrated the emergence of the compact Polyakov loop variables and of the singular field configurations which appear whenever the compact variables reach the border of the fundamental domain. Compactness of some of the variables  and the presence of singular  gauge fields not  suppressed by an infinite action, constitute  characteristic properties of the gauge fixed non-Abelian theory. 

In the following sections we will discuss to what extent the non-trivial phenomena of QCD may be associated with these properties of the gauge fixed theory. In this endeavor the Polyakov loop variables will play a central role. It will be an important asset that, in axial gauge, these variables which serve as order parameters occur as elementary rather than composite fields. Therefore in the absence of a viable approximation scheme, the known properties of QCD can in turn be used to deduce dynamical properties of these variables.

The central phenomenon of Yang-Mills theories is confinement. Polyakov loops are objects whose dynamics is intimately linked to this phenomenon. The spectrum of the Polyakov loops - with a spatial direction compact - reflects directly presence or absence of confinement.
We consider the correlation function of two Polyakov loops separated in Euclidean time. For large separation, this correlation function is given by 
$$ \lim_{t \rightarrow \infty}\langle 0|\mbox{tr}P(t_{E})\mbox{tr}P(0)|0 \rangle = c e^{-E_{-}t_{E}},$$
where, after Wick rotation, we  have made the following choice of coordinates  
$$x_{\perp} = (t_{E},0,0) .$$
 $E_{-}$ denotes the energy of the lowest state, which can be reached in applying the Polyakov loop operator with negative Z-parity to the ground state.
On the other hand, after a further rotation in the Euclidean, time and 3-axis can, up to a sign, be interchanged. In this operation the value of the correlation function does not change; it however now acquires a different interpretation. The correlator 
$$ \lim_{t \rightarrow \infty}\langle 0|\mbox{tr}P(x=t_{E})\mbox{tr}P(0)|0\rangle = e^{-LV(x)}$$ 
describes the same system at finite temperature with $T=1/L$ and the Polyakov loop correlator (corresponding to compact Euclidean time), as is well known~\cite{Svetitsky86},  is given by the free energy of a pair of static charges. In the confined phase
$$ 1/L = T < T_{c}, \quad V(r)=\sigma r$$
the interaction energy is linearly rising with the slope given by the string constant and therefore we conclude 
\beq
\label{gap-}
 E_{-} = \sigma L,\quad \mbox{if}\,\,\, L > L_{c} = 1/T_{c}.  
\eeq 
The lowest energy of states which can be excited by the Polyakov loop operator therefore increases linearly with the extension of the system. In particular, the ground state cannot contribute which is guaranteed provided the ground state is symmetric under center reflections (cf.\ Eq.(\ref{trpo})) 
$$\langle 0|P(x_{\perp})|0\rangle = 0, \quad \,\, \mbox{if}\,\, Z |0\rangle = |0\rangle . $$ 
In the deconfined phase we expect Debye screening to take place giving rise to an interaction energy of static charges
$$V_{D} \sim \frac{1}{r^{2}} e^{-m_{D}r} .$$
Following the same line of arguments this implies that, for extensions below the critical value,  the ground state breaks the center symmetry and  the excited states which can be reached by $P(x_{\perp})$ exhibit a gap $E_{-}$
$$L<L_{c}: \quad Z |0\rangle \neq |0\rangle \quad E_{-}=m_{D}.$$ 
A further characterization of the confined phase and the dynamics of the variables $a_{3}$ can be obtained through a discussion of  adjoint
Polyakov loops. The adjoint Polyakov loop is defined with the matrices
$T^{a}$ of the adjoint representation as
\beqs
 P_{\rm ad}(x_{\perp}) = \frac{1}{3} \mbox{tr}\,{\mbox P}
\exp \left(ig\int_{0}^{L}
dz A_3^{a}(x_{\perp},z)  T^{a}\right).
\eeqs
After gauge fixing also the adjoint loops are given in terms of the variables  $a_{3}$ (cf.\ Eqs.(\ref{axgc},\ref{shift}))
\beq
\label{adpl2}
 P_{\rm ad}(x_{\perp})=\frac{1}{3} \left(1-2\cos gLa_{3}(x_{\perp})
\right)\ .
\eeq
Adjoint charges can be screened by gluons; thus for sufficiently large distances, the interaction energy must tend exponentially to a constant with the exponential slope determined by the lowest glueball mass $m_{gb}$.
This implies a non-vanishing ground state expectation value  
$$\langle 0| \left(1-2\cos gLa_{3}(x_{\perp})\right)|\,0 \rangle \neq 0 \quad $$
and a gap $E_{+}$ of states which which can be excited by $ P_{\rm ad}$   
\beq
\label{gap+}
E_{+} = m_{gb} .  
\eeq
Being not forbidden by  symmetry requirements,  a non-vanishing vacuum expectation value is natural.
These facts concerning the dynamics of the variables $a_{3}(x_{\perp})$ strongly suggest the following properties of the spectrum of states which can be reached by applying operators built from these fields. In the confined phase, the ground state of the system is even under center reflections.  The spectra of excited states depend crucially on the $Z$-parity. States with positive $Z$-parity are physical, ``hadronic'' states, i.e.\ in pure Yang Mills theories they describe e.g.\ glueballs (with vanishing 3 component of the momentum $p_{3}$). States with negative $Z$-parity and vanishing  $p_{3}$ exhibit a gap which becomes infinite with infinite extension of the system. In other words in this limit these states are frozen and we may expect this to be the case also for states with  $p_{3}\ne 0 $ but significantly smaller than $1/L_{c}$. In the confined phase  no gauge-invariant operator can connect  states belonging to the 2 different sectors  due to their different $Z$-parity. The presence of quarks will substantially change this picture. With decreasing extension, the excitation energy of the $Z$-odd states decreases and when approaching the critical extension at which the confinement-deconfinement phase transition takes place, the gap in this sector is of the order of the glueball mass, which at infinite extension characterizes the gap in the $Z$-even sector. More quantitatively we know from lattice calculations~\cite{GGHK94,FIHK93} that in SU(3) with $T_{c} \approx 220$ MeV, the values of the lowest glueball mass and the gap in the Z-odd sector at $T\approx  170\,$MeV are
$$m_{gb}(170\,\mbox{MeV}) \approx 770\, \mbox{MeV} ,\quad \sigma(170\,\mbox{MeV})/T_{c}\approx 670 \,\mbox{MeV} .$$  
Thus at the phase transition the continuously decreasing gap suddenly vanishes together with the string tension and the $Z$-odd states become available and contribute to the thermodynamic quantities such as pressure or energy density. At the same time the glueballs in the $Z$-even  sectors disappear. With the ground state breaking the center symmetry, the two classes of states are now coupled and are therefore in thermodynamic equilibrium. Thus the confinement-deconfinement phase transition is not just a melting of the glueballs. Rather at the transition a whole sector of the Hilbert space, completely decoupled below the phase transition and not accessible to any physical observable, joins the physical states in the center-symmetry breaking plasma.       

\subsection{Theoretical Approaches to Polyakov Loop Dynamics}        

After having described the most prominent properties of the dynamics of the Polyakov loop variables we now turn to attempts to provide  theoretical understanding of some of these gross features of QCD. On the basis of the  expression (\ref{pm0a}) for the generating functional we will describe  a hierarchy of approximations in the evaluation of $Z$ with increasing complexity.  It is clear from the outset that, even for modest success, certain non-perturbative elements have to be incorporated.

The QCD generating functional in the naive axial (or temporal) gauge is obtained if only the sector without singularities is kept and the dependence on the eigenvalues of the Polyakov loops is disregarded. As a consequence of these approximations, the generating functional becomes actually ill-defined as has been noticed early by Schwinger~\cite{Schwinger}. Definition of propagators requires certain ``i$\epsilon$'' prescriptions. In the course of the approximations, the center symmetry got lost.

Still, keeping the zero singularity sector only one might proceed by accounting
for the dependence of $Z$ on $a_{3}$. The simplest form of these dynamics
results, if these variables are treated as Gaussian variables, i.e.\ if the
non-flat measure of Eq.(\ref{FE8p}) is replaced by the flat one 
\beqs D\left[a_{3}\right] \approx \prod_{x_{\perp}} da_{3}\left(x_{\perp}\right) .
\eeqs
In this way, one
effectively treats the Polyakov loop eigenvalues as the zero modes in QED. It
is therefore not surprising that the center-symmetry is lost again and and a perturbative picture emerges with the phenomenon of Debye
screening as the leading dynamical correction to the description of QCD as a system of non-interacting gluons~\cite{Weiss}.

As we now will discuss in more detail, first characteristic properties of QCD are encountered if, still disregarding  singular 
field configurations, the non-flat measure of the Polyakov loop variables is
properly taken into account. In particular, the perturbative phase reached in
this way will be seen to be center-symmetric. In the last section  we will address the possible role of the singular field configurations in axial gauge QCD.

A crucial element in the following discussion will be the compact, i.e.\ non-Gaussian
nature of the Polyakov loop variables $a_{3}(x_{\perp})$. We have seen that the appearance of compact variables is common to most of the formulations of QCD in terms of unconstrained variables.  To study the consequences we use the canonical formulation and disregard in a first step the coupling of  Polyakov loop variables to the other degrees of freedom \cite{LeMT95}. In the absence of such couplings, the  Hamiltonian of the Polyakov loop variables reads (cf. Eq.(\ref{hamaxg})) 
\beqs
h=\int d^{2}
x_{\perp}\left[-\frac{1}{2L}\frac{\delta^{2}}{\delta a_3({\bf x}_{\bot})^{2}} +\frac{L}{
2}\left(\mbox{\boldmath$\nabla$}a_3({\bf x}_{\bot})\right)^2\right] \ .
\eeqs
and if space is discretized (lattice spacing $\ell$)
\beq
h= - \frac{g^{2}L}{8 \ell^{2}} \sum_{\vi}
\frac{\partial^{2}}{\partial \tilde{a}_{3}(\vi) ^{2}} 
+ \frac{2}{g^{2}L}
\sum_{\vi,\mbox{\boldmath$\scriptstyle\delta$}}\ \left(\tilde{a}_{3}(\vi +\mbox{\boldmath$\delta$})-
\tilde{a}_{3}(\vi) \right)^{2}=h^{e}+h^{m}\ ,
\label{19}
\eeq
where $\mbox{\boldmath$\delta$}=(\ell,0),(0,\ell)$ denote the fundamental vectors of the lattice and the dynamical variables have been rescaled
\beq
\label{resc}
 \tilde{a}_{3}(x_{\perp}) = gLa_{3}(x_{\perp})/2 \ . 
\eeq
 For weak coupling to other degrees of freedom, the Hamiltonian (\ref{19}) describes both, photons in QED and  Polyakov loop variables in QCD. It however acts on wave functions  belonging to different spaces. In QCD  the compact nature imposes the boundary condition (cf.\ Eq.(\ref{abc},\ref{shift}))
\beq
\hat{\Psi}[\tilde{a}_{3}]=0 \quad \mbox{whenever} \quad \tilde{a}_{3}(\vi)=
\pm \frac{\pi}{2} \quad \mbox{for some}\  \vi \ .
\label{21}
\eeq
In order to display the non-trivial dynamics described by the
Hamiltonian $h$ in conjunction with the constraint of
Eq.(\ref{21}) we consider first the case of electrodynamics where no such constraint is present. 
As is well known, by discrete Fourier transformation, the elementary excitation can be determined and the following dispersion relation  
 for the (lattice) photons,
\beq
\omega_{{\bf k}}^{2}=\frac{4}{\ell ^{2}}\sum_{\mbox{\boldmath$\scriptstyle\delta$}}
\sin ^{2}\frac{(\mbox{\boldmath$\delta$}{\bf k})^{2}}{4} \quad {\bf k} = \frac{2\pi}{L}(n_{1},n_{2}) 
\label{disc}
\eeq
is obtained, with the standard continuum limit
\beqs
\omega_{{\bf k}}^{2} \rightarrow {\bf k}^{2} \ .
\eeqs
The ground state wavefunctional 
\beq
\label{gstwved}
\hat{\Psi}[a_{3}] = \exp \left\{-\frac{1}{2} \int d^{2}x_{\perp}d^{2}y_{\perp} a_{3}(\vi)K(\vi,\vj)a_{3}(\vj)\right\}   
\eeq
expresses by its non-locality 
$$K(\vi,\vj) = L \langle \vi|\sqrt{-\Delta_{\perp}}| \vj\rangle$$
strong correlations in the system.
While  electric and  magnetic field energy contribute equally  in the normal modes of QED,  in QCD as a consequence of the  boundary condition (\ref{comvar})
$$\left(\tilde{a}_{3}(\vi +\mbox{\boldmath$\delta$})-
\tilde{a}_{3}(\vi) \right)^{2} \le \pi^{2}$$
 the magnetic field energy $h^{m}$ becomes negligible in the continuum limit ($L/\ell \rightarrow \infty$). 
Comparable contributions from electric and magnetic field energy could result  in the continuum limit  only if we assume linear instead of logarithmic running of the coupling constant
$$g^{2}\sim \ell, \quad \ell \rightarrow 0 .$$
Dominance of the electric field energy yields the reduced ground state wavefunctional
\beq
\hat{\Psi}_{0}\,[\,\tilde{a}_{3}] = \prod_{\vi} 
\left[ \left(
\frac{2}{\pi}\right)^{1/2}
 \cos \left( \tilde{a}_{3}(\vi) \right) \right]   .
\label{35}
\eeq
States of  lowest excitation   energy are obtained by exciting a
degree of freedom at one site ${\tilde{\bf x}}_{\perp}$ into its first excited
state; this is achieved by replacing in the ground state wave functional 
\beqs
 \cos \left( \tilde{a}_{3}({\bf \tilde{x}}_\perp ) \right) \rightarrow \sin \left(2 \tilde{a}_{3}( {\bf \tilde{x}}_\perp ) \right) 
\eeqs
and an  excitation energy
\beq
\Delta E = \frac{3}{8} \frac{g^{2} L}{\ell ^{2}} \ 
\label{38}
\eeq
results. Such excited states can occur at any site and therefore these states are highly degenerate.  This degeneracy is lifted by the magnetic coupling. A perturbative evaluation yields a ``band'' of of excited states characterized by the discrete momenta (cf.~(\ref{disc})) and with excitation energies
\beqs
\Delta E_{{\bf k}} = \frac{3}{8} \frac{g^{2} L}{\ell ^{2}}+\frac{64}{9\pi^{2}}
\frac{1}{g^{2}L}\left[\sum_{\mbox{\boldmath$\scriptstyle\delta$}}\sin^2 \frac{(\mbox{\boldmath$\delta$}
{\bf k})^{2}}{4} - 1 \right]
\eeqs
The qualitative differences in the structure
of the ground states of QED and QCD respectively (Eqs.(\ref{gstwved},\ref{35})) imply very different properties of the corresponding elementary
excitations. Built on the highly correlated QED ground
state  the photons appear as  collective  excitations with excitation
energies vanishing in the  long-wavelength limit. In QCD,
the elementary excitations are localized in configuration space and
are due to formation of non-vanishing electric flux. In the absence of couplings to the other degrees of freedom, the
chromoelectric fields formed in the elementary excitations with lowest
excitation energy are located on just one transverse lattice site and the flux tube is infinitely thin. It winds around the compact 3 direction and thereby gives rise to an excitation energy which increases linearly with the extension  i.e.\ in this limit
$$E_{-} = \sigma_{stc}L ,$$
with the string tension given by the strong coupling lattice result~\cite{KOSU}
$$\sigma_{stc}=\frac{3g^{2}}{8\ell^{2}} .$$ 
 The ground state of the system
is, in the strong coupling limit, an eigenstate of the electric
field operator. 
In QED with photons propagating in the $1-2$ plane described by Gaussian variables $a_{3}({\bf x}_{\perp})$ such states are not normalizable and would entail infinitely
large fluctuations in the magnetic field energy. Thus the
structure of the vacuum concerning these $x_{3}$-independent
fields is very different in the Abelian and non-Abelian theory.
In QED  the virial theorem yields the standard result: magnetic and electric fields
contribute equally to each normal mode.
In axial gauge QCD the Jacobian invalidates
equipartition. Chromoelectric $x_{3}$-independent
fields are absent; the square of the reduced ground state wave functional (\ref{35}) is nothing else than the Jacobian (\ref{FE8p}) and thus the corresponding full wave-functional is  constant. In turn, the fluctuations in the magnetic
field at different lattice sites are not correlated and the ground state
energy is due exclusively to these uncorrelated magnetic field fluctuations.
Therefore the corresponding contributions to  the ``gluon condensate" are dominated by the magnetic
field 
\beqs
\delta \langle 0| {\bf E}^2-{\bf B}^2 |\,0\rangle = -\delta \langle 0|\, {\bf B}^2\, |0\rangle  \ .
\eeqs
With the electric flux quantized and resulting in the gap (\ref{38}) in the spectrum, this magnetic field dominated ground state of the Polyakov loop variables exhibits a dual Meissner effect, i.e.\ it resists penetration of $x_{3}$ independent electric fields pointing in the spatial $3$ direction into the medium. This is not a result of a condensation of monopoles but a consequence of the compactness of the relevant gauge fields. 
It is interesting that even in the crude approximation of keeping only
one particular kind of gluonic degrees of freedom, this model
displays features which are reminiscent of the phenomenology of the
``magnetic QCD vacuum"~\cite{NACH97}.
The compact nature of the Polyakov loop variables appears to reproduce the most striking  feature of the spectrum in the $Z$-parity odd sector of the theory (cf.\ Eq.(\ref{38})), the linear increase of the gap with the extension. It thus displays certain confinement like properties, it gives rise to a magnetic vacuum and exhibits the dual Meissner effect. Quantitatively, the description of these variables as infinitely thin flux tubes and the deduced strong coupling value of the string constant is  unrealistic; accounting for the coupling to the other degrees of freedom  might be expected to improve these results. On the other hand  at this level of the development,  the formalism fails in not  displaying any significant differences in the dynamics of confined Z-odd and ``hadronic'' Z-even states.

\subsection{Perturbative Results for the Confined Phase}

The following studies concerning  the effects of the coupling of the Polyakov loops to the other gluonic variables will be performed on the basis of the generating functional (\ref{pm0a}). In particular we will be interested in the properties of the Polyakov loop correlation functions. For their evaluation, we observe that the (in the continuum limit) infinitely large gap (cf.\ Eq.(\ref{38})) prohibits the Polyakov loop variables to propagate. Indeed our above results are easily translated into properties of the relevant correlation functions. In  the absence
of coupling to the other degrees of freedom the generating functional written for discretized space-time and in terms of the rescaled variables (\ref{resc}), is in the Euclidean, given by
\beas Z_{0}\!\! & = & \! \int d\left[a_{3}\right] 
\exp \left\{-1/2 \int d^{4}x(\partial_{\mu} a_{3}(x_{\perp}))^{2}\right\}
 \\ & = & \! \int\limits_{-\frac{\pi}{2}}^{\frac{\pi}{2}} \prod_{x_{\perp}}
d\tilde{a}_{3}\left(x_{\perp}
\right) \cos^{2}\tilde{a}_{3}\left(x_{\perp}\right)\exp \Bigl\{\frac{-2\ell} {g^{2}L}
\sum_{y_{\perp},\delta_{\perp}}(\tilde{a}_{3}( y_{\perp}+\delta_{\perp})-
\tilde{a}_{3}( y_{\perp}))^{2}\Bigr\} \ .
\nonumber
\eeas
 In the continuum limit
\beqs
\frac{\ell}{g^{2}L} \sim \frac{\ell}{L} \ln \frac{L}{\ell}
\rightarrow 0 \ ,
\eeqs and therefore the nearest neighbor interaction generated by the Abelian
field energy of the Polyakov loop variables is negligible. As a consequence, in the
absence of coupling to other degrees of freedom, Polyakov loops are ultralocal, they do not propagate,
\beqs
\langle \Omega|T\left( a_{3} \left(x_{\perp}\right) a_{3} \left(0\right)
\right)|\Omega\rangle  \sim \Bigl(\frac{\ell}{g^{2}L}\Bigr) ^{x_{\perp}/\ell} \! \!
\rightarrow
\delta^{3}
\left(x_{\perp}\right) \ .
\eeqs Propagation of excitations induced by $a_{3} (x_{\perp})$ can only arise by coupling to the other microscopic degrees of freedom. Ultralocality permits the Polyakov loop variables
$a_{3} $ to be integrated out and the following effective action results
\beq 
S_{\rm eff} \left[A_{\mu}\right] = S_{\rm YM} \left[A_{\mu},
A_{3}=0\right] + S_{\rm gf} \Bigl[ \int_{0}^{L} dz\, A_{\mu}^{3}\Bigr] +  M^{2}
 \int d^{4}x\,\, \Phi_{\mu}^{\dagger}(x) \Phi^{\mu}(x) .
\label{FE11}
\eeq 
The Polyakov loop variables have left their
signature in the  geometrical mass term of the charged gluons (cf.\ Eq.(\ref{gf12a}))
\beq
 M^{2}=(\pi^{2}/3-2)/L^{2}
\label{FE12}
\eeq
 and in the antiperiodic  boundary conditions  (Eq. (\ref{cs3c})). The
neutral gluons remain massless and periodic. The antiperiodic boundary conditions
reflect the mean value of the Polyakov loop variables, the geometrical mass their
fluctuations; notice that in both of these corrections, the coupling constant has dropped
out. We emphasize that periodic boundary conditions for the gluon fields are imposed in
the original expression (\ref{gfax1}) of the generating functional. The antiperiodic boundary
conditions implicit to the definition of the generating functional in (\ref{pm0a}) account for the appearance of Aharonov-Bohm fluxes in the
elimination of the Polyakov loop variables. Periodic charged gluon fields may be continued to be used if
the differential operator
$\partial_{3}$ is replaced by
\beq
\partial_{3} \rightarrow \partial_{3}+\frac{i\pi}{2 L} \left[\tau_{3}\right. , \qquad .
\label{FE13a}
\eeq
 As for a quantum mechanical particle on a circle, such a magnetic flux
is technically most easily  accounted for by an appropriate  change in boundary
conditions -- without changing the original periodicity requirements. With regard to
the rather unexpected physical consequences, the space-time independence of this flux
is important, since it induces global changes in the theory. These global changes are
missed if Polyakov loops  are treated as Gaussian variables. 

The role of the order parameter is taken over by the neutral color current in
3-direction $u\left(x_{\bot}\right)$ which is generated by the 3-gluon interaction 
\beq u \left(x_{\perp}\right) = i \int_{0}^{L} dx_{3}\,  \Phi^{\dagger}_{\mu}
\left(x\right)\stackrel{\leftrightarrow}{\partial}_{3}  \Phi^{\mu}
\left(x\right) \ .
\label{I54}
\eeq
 This composite field is odd under center reflections (cf.(\ref{gf12b}))
\beqs Z: \quad u(x_{\perp}) \rightarrow -u(x_{\perp}) . 
\eeqs It determines the vacuum expectation value of the Polyakov loops
\beqs
\langle \Omega| P \left(x_{\perp}\right)|\Omega \rangle \quad
\propto \quad \langle \Omega| u \left(x_{\perp}\right)|\Omega \rangle
\eeqs and the corresponding correlation function
\beqs
\langle \Omega| T\left[P \left(x_{\perp}\right) P \left(0\right)\right] |\Omega
\rangle  \quad  \propto \quad \langle \Omega| T\left[u
\left(x_{\perp}\right) u \left(0\right)\right] |\Omega \rangle ,
\eeqs
 which in turn yields the static quark-antiquark interaction energy~\cite{Svetitsky86}. Up to an irrelevant factor we have after rotation to the Euclidean ($r= |
x^{E}_{\perp} |$)
\beqs 
\exp{\left\{-LV\left(r \right)\right\}}= \langle \Omega|T \left[u ( x^{E}_{\perp})
u\left(0\right)\right]| \Omega\rangle ,
\eeqs 
i.e.\ the static quark-antiquark potential is given directly by (the
$a=b=3$, $\mu = \nu = 3$ component of)  the  vacuum polarization tensor $\Pi_{\mu
\nu}^{ab} $ and not by the zero mass propagator with corresponding self-energy
insertions as obtained in the standard Gaussian treatment. 

Similarly for evaluation of  the adjoint Polyakov loop correlator (cf.\ Eq.(\ref{adpl2})) we introduce  the composite field $v(x_{\perp})$
\beq
\label{defv}
  v \left(x_{\perp}\right) =  \int_{0}^{L} dx_{3}\,  \Phi^{\dagger}_{\mu}
\left(x\right) \Phi^{\mu}
\left(x\right) .
\eeq
Under center reflections $v$ is invariant
\beqs Z: \quad v(x_{\perp}) \rightarrow v(x_{\perp}) ,   
\eeqs 
and expectation value and correlator of $P_{\mathrm{ad}}(x_{\perp})$ are given by
\beq
\langle \Omega |P_{\rm ad}|\Omega \rangle  \sim
\langle \Omega|  v |\Omega \rangle 
\ \ , \ \  
\langle \Omega |T  \left[  P_{\rm ad}(x_{\bot})  P_{\rm ad}(0)
\right] |\Omega
\rangle
\sim \langle \Omega |T  \left[ v(x_{\bot}) v(0) \right] |\Omega \rangle \ .
\label{SP27d}
\eeq

\begin{figure}
\noindent
\begin{center}
\begin{minipage}[b]{0.75\linewidth}
\epsfig{file=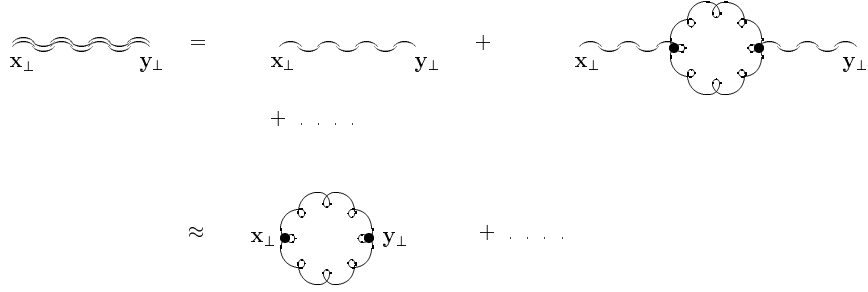, width=\linewidth}
\caption{Ultralocality of the Polyakov loops. Wavy Lines denote Polyakov loop propagators, curly lines charged gluon propagators.} \label{Fig2}
\end{minipage} \end{center}
\end{figure}

The system described by the effective action (\ref{FE11}) exhibits remarkable
properties already at the perturbative level. Most importantly the center symmetry is
realized in the perturbative vacuum, i.e.\ in the ground state obtained by dropping all
the terms containing the coupling constant $g$. Geometrical mass (Eq.(\ref{FE12})) and
Aharonov-Bohm flux (Eq.(\ref{FE13a})) are not affected by such a perturbative
treatment. The perturbative ground-state is even under charge conjugation and the
expectation value of the Polyakov loop vanishes 
\beqs
\langle\Omega_{\rm pt}| P \left(x_{\perp}\right)|\Omega_{\rm pt}
\rangle = 0 ,  \ 
\eeqs indicating an infinite free energy of a static quark. Indeed a perturbative
analysis of the correlation function confirms this expectation.  
The novel aspects of perturbation theory in the  center symmetric phase are related to
the ultralocality property of the Polyakov loops. As a consequence of
ultralocality, the Polyakov loop correlator is given by a one particle
 irreducible 2-point function, i.e.\ by the vacuum polarization  rather than by
the one particle reducible Green function of standard Gaussian variables. This is illustrated in Fig.~\ref{Fig2}.
Thus the interaction energy of static charges is determined by the color neutral, $\mu=\nu=3$ component of the vacuum polarization tensor 
\beqs
V(r)=-\frac{1}{L} \ln \int \frac{d^3 p}{(2\pi)^3}
\Pi^{33}(ip_0,p_1,p_2)e^{i \vec{p}\vec{r}} .
\eeqs
In one loop approximation (Fig.~\ref{Fig2}) $\Pi^{33}$ is given  by
\beqs
\Pi^{33}(p) = \frac{2 g^2}{L} \sum_{q_3} q_3^2
\int \frac{d^3q}{(2\pi)^3} \sum_{a,b \neq 3}D_{\mu \nu}^{ab}(q,
q_3) D^{ab, \mu \nu} (p-q,q_3)             ,
\eeqs 
with the  charged gluon propagator 
\beqs
D_{\mu \nu}^{ab}(p,p_3) = \frac{\delta^{ab}}
{p^2-p_3^2-M^2+i\epsilon}
\left[-g_{\mu \nu} + \frac{p_{\mu} p_{\nu}}{p_3^2+M^2} \right] \ , \quad 
 a\ne 3,\quad p_3= \frac{2\pi}{L} \left(n +\frac{1}{2}\right) \ .
\eeqs
Using dimensional regularization in the evaluation of $\Pi^{33}$, ultraviolet divergencies occur only in the final sum over Matsubara wave-numbers. They are most conveniently regularized by a heat kernel method.  Due to the antiperiodic boundary conditions of the charged gluon fields no infrared divergencies occur.
The final result for $\Pi^{33}$ yields the following asymptotic behavior of $V$
\beqs
V(r) \approx  \sigma_{\rm pt} r + \frac{2}{L}
\ln (r) + \mbox{O}(1)
\ , \qquad r/L \gg 1 \ .
\eeqs
with the perturbative value of the string constant
\beqs
 \sigma_{\rm pt} = \frac{2}{L}\sqrt{M^{2}+\frac{\pi^{2}}{L^{2}}}\,\, .  
\eeqs
The leading term in $V$ is independent of the coupling constant. It is determined kinematically. At large separations, the dominant contributions to $V$ arise from the singularities of the polarization tensor (cf.\ Fig.~\ref{Fig2}) closest to the real axis. They describe production of 2 charged gluons of mass $M$ and minimal momentum ($\pi/L$). The perturbative string tension is up to the factor $1/L$ the threshold energy for this production process. With this argument we immediately can deduce the interaction energy of adjoint charges determined by the correlation function of the composite fields $v(x_{\perp})$ (cf.\ Eqs.(\ref{defv},\ref{SP27d})). Once more the large distance behavior of $V_{\rm ad}$ is dominated by the threshold of two (charged) gluon production; in this case however also the ground state contributes. These considerations determine  the asymptotic behavior of $ V_{\rm ad}$ up to a constant
$$e^{-L V_{\rm ad}(r)}= \langle \Omega_{\rm pt} | P_{\rm ad}|\Omega_{\rm pt} \rangle ^{2} + c \frac{1}{r} e^{-2\sigma_{\rm pt}r} $$
and thus
$$V_{\rm ad}(r) \sim \frac{1}{r} e^{-2\sigma_{\rm pt}r} .$$ 
The characteristic difference in the asymptotic behavior of  the interaction energies of fundamental and adjoint charges respectively is due to the ultralocality and ultimately has its origin in the compactness of the Polyakov loop variables.\\
As a final application, we discuss the effect of dynamical quarks on the interaction energy of static fundamental charges. Dynamical quarks give rise to an additional contribution to the order parameter field $u(x_{\perp})$ in Eq.(\ref{I54}))
\beqs u \left(x_{\perp}\right) =  \int_{0}^{L} dx_{3}\,\left\{ i \Phi^{\dagger}_{\mu}
\left(x\right)\stackrel{\leftrightarrow}{\partial}_{3}  \Phi^{\mu}
\left(x\right) -\bar{\psi} \left(x\right)  \frac{\tau_{3}}{2}\gamma_{3}
\psi\left(x\right)\right\} \ .
\eeqs
The anti-periodic boundary conditions of the quark fields have been changed in the field redefinition  Eq.(\ref{shift}) to
\beq
\label{qpbc}
\psi(x_{\perp}, x_{3} =L )= e^{-i\pi\scriptstyle\tau_{3}/2}\psi(x_{\perp}, x_{3} =0 ).  
\eeq
As for the charged gluons, this change in boundary conditions 
accounts for the interaction with the Aharonov--Bohm fluxes generated by the
ultralocal Polyakov loop variables. As is clear from our above discussion of the interaction energy of adjoint charges, unlimited increase in the  interaction energy with increasing  separation of the static charges  will avoided only if  $u \left(x_{\perp}\right)$ develops a non-vanishing vacuum expectation value
\beqs
\exp{\left\{-L V\left(r \right)\right\}} \approx  \left|
\langle \Omega_{\rm pt}|u\left(0\right)|\Omega_{\rm pt}\rangle  \right|^{2}.
\eeqs
Applying again perturbation theory we have
\beqs
\frac{1}{L}  \langle \Omega_{\rm pt}|u\left(x_{\perp}\right)|\Omega_{\rm pt}\rangle  =  - \langle \Omega_{\rm pt}|
\bar{\psi} \left(x\right)  \frac{\tau_{3}}{2}\gamma_{3}
\psi\left(x\right)|\Omega_{\rm pt}\rangle = \frac{1}{L}\sum_{k_{3}} \int
\frac{d^{2}k}{(2\pi)^{2}} \frac{2 k_{3}}{E_{k}}  .
\eeqs
The divergent expression for the vacuum expectation value
can be evaluated by performing the integral over transverse momenta in dimensional
regularization and yields the result 
\beqs
\langle \Omega_{\rm pt}|u\left(x_{\perp}\right)|\Omega_{\rm pt}\rangle= \frac{2 m^{2}}{\pi^{2} } \sum_{n=1}^{\infty}
\frac{\sin 2\pi
n\alpha}{n} K_{2}(nmL) \ .
\eeqs
The parameter $\alpha$ characterizes the boundary conditions; $\alpha = 0,1/2$ denote periodic and anti-periodic boundary conditions respectively. For these values the order parameter $u$ does not develop a vacuum expectation value. The value $\alpha = 1/4$ denotes the quasi-periodic b.c. of  Eq.(\ref{qpbc}) and yields a non-vanishing expectation value.
For quark masses $mL \gg 1$
$$ \langle \Omega_{\rm pt}|u\left(x_{\perp}\right)|\Omega_{\rm pt}\rangle \sim e^{-mL}$$
and thus for asymptotic separations, the interaction energy of two static charges
reaches the constant value
\beq
\label{CE6}
V(r) \approx 2m \ .
\eeq

It is remarkable that this correct form of the asymptotics of $V$ appears within perturbation theory. In standard perturbation theory quark loops  yield as
in QED an Uehling type correction~\cite{Uehling}
to the Coulomb interaction $\delta V \propto
g^{4} \exp{(-2mr)}/r$. Perturbation theory in the center symmetric phase on the
other hand yields the coupling constant independent result (\ref{CE6}). Thus the
mechanism of string breaking by pair formation is apparently present in the
center symmetric phase already at the perturbative level. The calculation
also displays the important role
of the modification of the boundary conditions, i.e.\ the role of the
Aharonov--Bohm
fluxes. The string breaking mechanism would not be present if dynamical quarks
satisfied standard
anti-periodic boundary conditions, nor would it arise if the boundary
conditions would not differentiate
between the two color states of the quarks
($\tau_{3}$ in the b.c.~(\ref{qpbc})).

\subsection{Polyakov Loops in the Plasma Phase}

In the last two sections we have identified the role of the Polyakov loop variables in the confined phase of QCD. We have displayed  the relevance of these variables for both a phenomenological characterization of the confined phase and for  development of a modified perturbative approach. We now will extend our investigations and will exhibit the particular role of these variables in the deconfined phase. The perturbative center symmetric phase with its signatures of confinement cannot be relevant for QCD at
extensions smaller than $L_c=1/T_{c}$ or temperatures beyond the critical temperature $T_{c}$.  Not only do we expect the center symmetry to be
 broken at small extensions but also dimensional reduction to
 QCD$_{2+1}$ to happen. Aharonov-Bohm fluxes  induce anti-periodic boundary conditions and therefore yield a decoupling of the charged gluons
 if dimensional reduction takes place in the
center symmetric phase (high temperature confining phase).   The small extension or high temperature limit of the center
symmetric phase is therefore QED$_{2+1}$. In comparison to the dimensional reduction to  QCD$_{2+1}$, the resulting values for energy density and pressure would be reduced for SU$(N)$ Yang-Mills theories by a factor $1/(N+1)$. This is in conflict with the results from lattice calculations.  In order to reach the correct high
temperature phase, the deconfinement phase-transition arising when compressing the
QCD vacuum, must be accompanied by screening of the Aharonov-Bohm fluxes and 
simultaneously by a weakening of the increase in the geometrical mass. In the
following we shall treat the
Aharonov-Bohm fluxes (Polyakov loop variables) as phenomenological space-time independent quantities  and as is appropriate for the deconfined phase, shift back the origin of these variables   
(cf.\ Eq.(\ref{shift})) 
\beqs
  \chi= gLa_{3} +\pi .
\eeqs
For simplicity we assume the geometrical mass $M$ (cf.\ Eq.(\ref{FE12})) to vanish in the
deconfined phase. 

\begin{figure}
\noindent
\begin{minipage}[b]{0.44\linewidth}
\epsfig{file=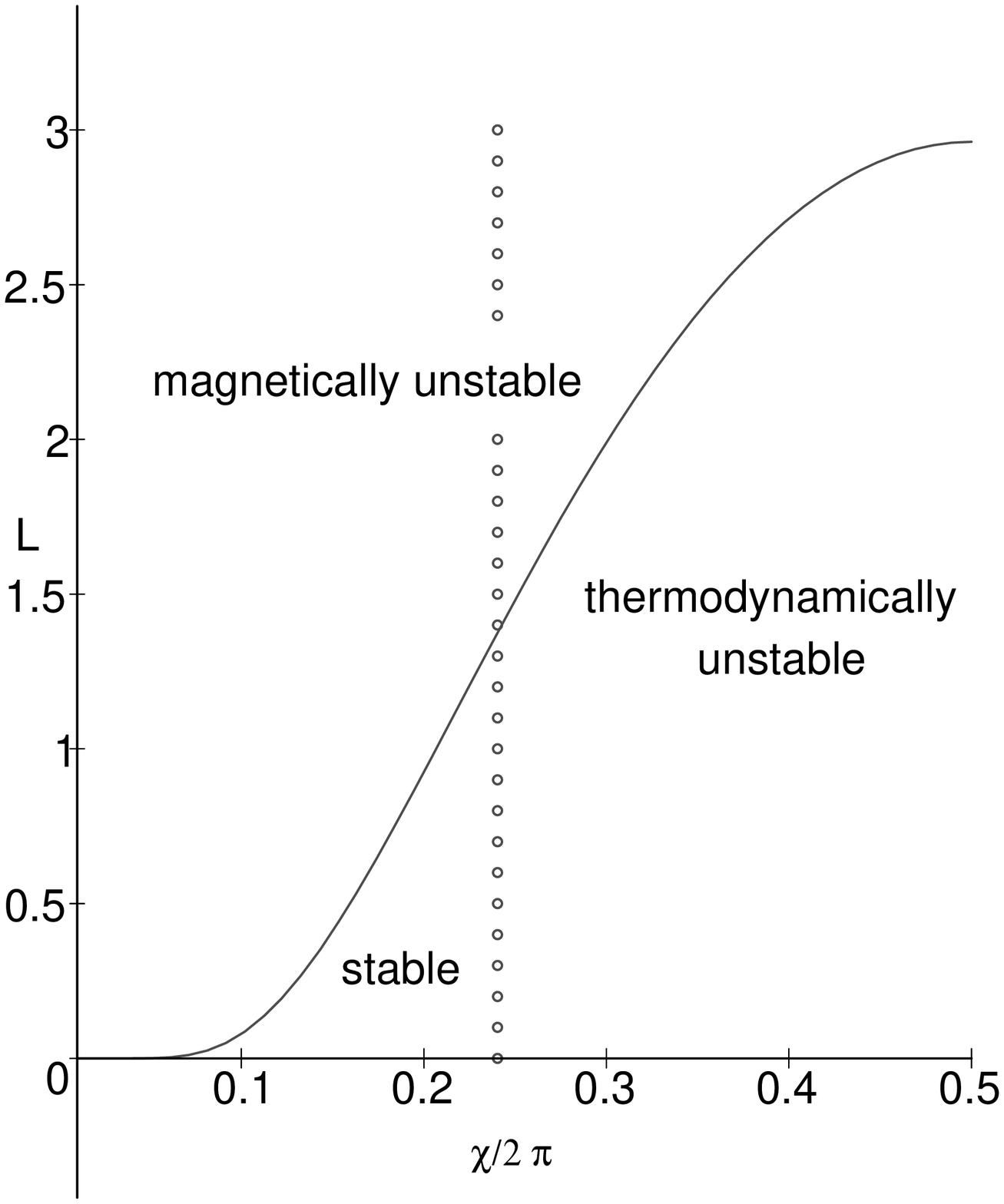,width=\linewidth}
\caption{Regions of stability and instability in the $(L,\chi)$ plane. To the right of the circles, thermodynamic instability; above the solid line, magnetic instability.} \label{Fig3}
\end{minipage}
\hfill
\begin{minipage}[b]{0.44\linewidth}
\epsfig{file=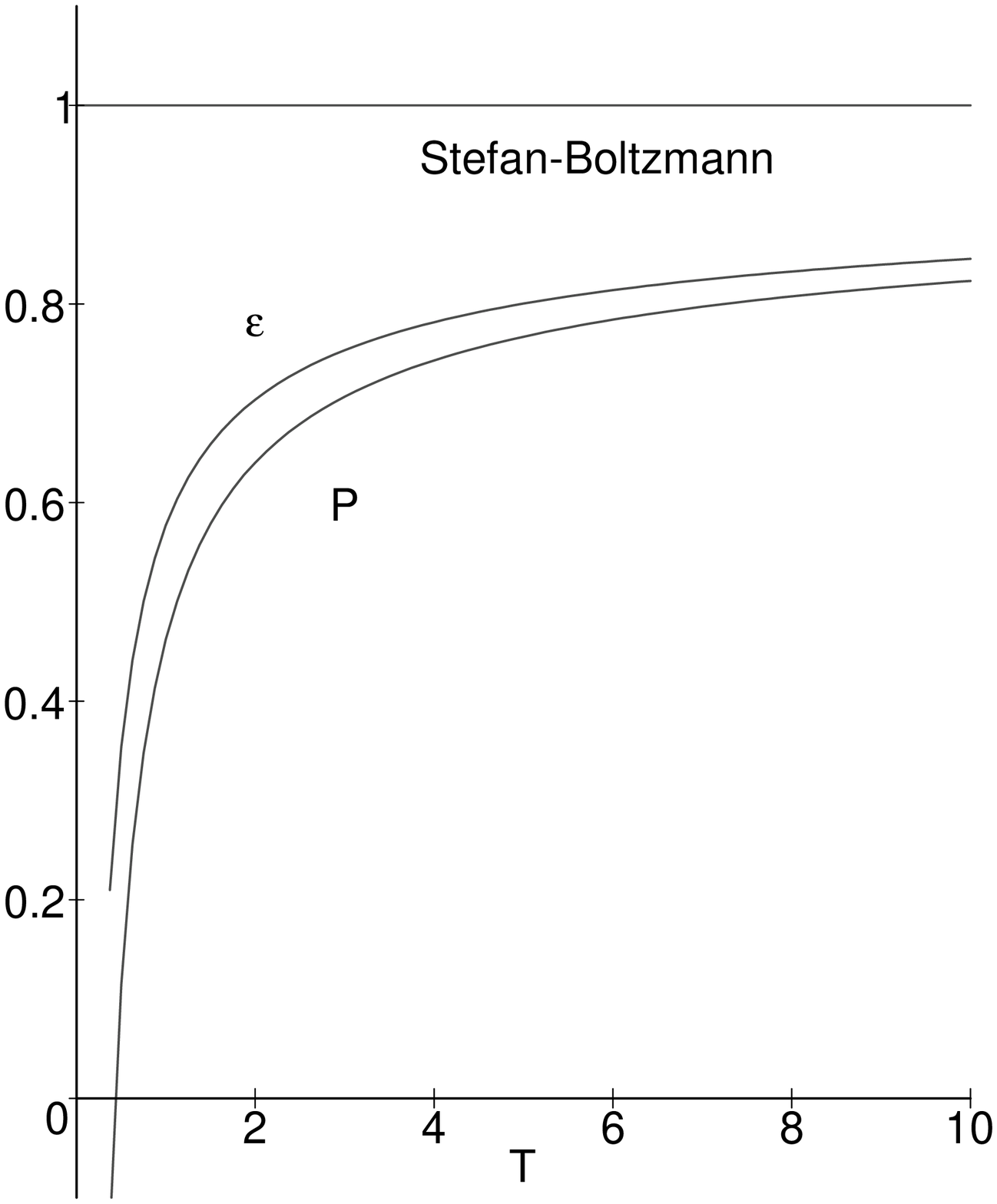,width=\linewidth}
\caption{Energy density and pressure normalized to Stefan-Boltzmann values vs. temperature in units of $\Lambda_{\rm MS}$.} \label{Fig4}
\end{minipage}
\end{figure}

The strength $\chi$ is limited by the requirement of
thermodynamic stability. By covariance, positive Casimir energy $\epsilon$ at finite extension
corresponds to a negative pressure of the corresponding system at
finite temperature. The expression for the Casimir energy of the charged gluons  
\beqs
\varepsilon(L,\chi) = - \frac{4}{\pi^2 L^4}
\sum_{n=1}^{\infty}\frac{\cos (n\chi)}{n^4}
 =  \frac{4\pi^2}{3L^4} B_4\left( \frac{\chi}{2\pi} \right),\quad
 B_{4}(x)=-\frac{1}{30} +x^{2}(1-x)^2
\eeqs
yields the following requirement for thermodynamic stability  
\beq
\label{thst}
 \chi\leq 1.51 . 
\eeq
Complete screening ($\chi = 0$) of the Aharonov-Bohm fluxes, compatible with thermodynamic
stability,  is
unlikely to take place in the deconfined phase. Such a system will exhibit, like perturbative QCD at zero
temperature, a  magnetic instability~\cite{SAV77}. Spontaneous formation
of magnetic fields on the other hand is prevented in the presence of
sufficiently strong Aharonov-Bohm fluxes. Calculation of the Casimir energy for gluons subject to a color-magnetic homogeneous background field 
\beqs
A^a_\mu|_{\rm bg} =  \delta^{a3} \delta_{\mu 1} x_2 H \ .
\eeqs
leads, for given extension $L$ or temperature $T$ to a lower limit
for $\chi$. Figure \ref{Fig3} displays the region of thermodynamic and magnetic
stability. Obviously, due to the requirement of magnetic stability, the
Stefan-Boltzmann limit (corresponding to $\chi=0$) cannot be reached for any
finite temperature. Identification of the Aharonov-Bohm flux with the minimal
allowed values sets upper limits to energy density and pressure. They are  shown in
Figure \ref{Fig4} and are reminiscent of lattice data~\cite{EKSM82} in their slow,
logarithmic approach to the Stefan-Boltzmann limit. The finite value of the
Aharonov-Bohm flux accounts for interactions present in the deconfined phase
and specifically gives rise to values of the interaction measure $\epsilon-3P$ which are in
qualitative agreement with lattice data. Other quantities like the Debye
screening mass are also affected by the non-vanishing Aharonov-Bohm 
flux. A one loop calculation 
\beqs
  m_{el}^2 = \frac{4 g^2}{L^2} B_2\left( \frac{\chi}{2\pi} \right), \quad
B_2(x)=1/6-x(1-x)  
\eeqs
reproduces qualitatively the approximate linear dependence of  $ m_{el}^{2}$ over a large
temperature regime as observed in
lattice calculations. Finally these results can also be used to estimate the
change in energy density $-\Delta \epsilon$ across the phase transition. In this phenomenological
treatment the phase  transition is accompanied by a change in strength of the
Aharonov-Bohm flux from the center symmetric value $\pi$ to a value in the stability region of Fig.~\ref{Fig4}. The lower bound is determined by  
Eq.(\ref{thst})
\beqs
 - \Delta \epsilon = \epsilon (L_{c},
  \chi=\pi)-\epsilon (L_{c}, \chi=1.51)  \leq \frac{7\pi^{2}}{180}\frac{1}{L_{c}^{4}} .
\eeqs
For deriving the upper bound  the critical temperature must be specified. With $T_{c} \approx 270$ MeV we obtain 
\beqs
0.38\,\frac{1}{L_{c}^{4}}\le - \Delta \epsilon  \le 0.53\,\frac{1}{L_{c}^{4}}.
\eeqs  
The lattice result~\cite{Engels} is within these limits
\beqs
\Delta \epsilon=  -0.45\frac{1}{L_{c}^{4}} \ .
\eeqs
It appears that to large extent, the dynamics in the deconfined phase is summarized by the dynamics of the Polyakov loop variables. The picture of increasing  screening of the Aharonov-Bohm fluxes with increasing temperature seems to catch the essential physics of the  thermodynamic quantities. It is remarkable that the requirement of magnetic stability which prohibits complete screening seems to determine the temperature dependence of the Aharonov-Bohm fluxes and thereby to simulate the non-perturbative dynamics  in a semiquantitative way.  

\subsection{Monopole Dynamics}

In the last sections we have investigated the dynamics of the Polyakov loop variables. We have described phenomenological implications and have
demonstrated the important role of the compactness of these variables in the structure of the theory. In particular, the compactness offers the
possibility of a perturbative approach to certain confinement related phenomena of QCD. For instance, when neglecting couplings to  other gluonic
degrees of freedom, the dynamics of  the Polyakov loops is determined completely by their compact nature and yields  a gap in the spectrum which
increases linearly with the extension of the system. This in turn implies a linearly rising interaction energy of static charges with a value of 
the string tension which becomes infinite in the continuum limit. If in a second step, gluonic couplings are taken into account perturbatively, the 
linear rise in the interaction energy persists with a value of the string tension which decreases with increasing extension of the system. Other 
qualitative features are properly described, such as the absence of confinement for adjoint charges or the string breaking mechanism induced by 
dynamical quarks.  The formalism however also indicates that realistic values of the string tension are not attainable in this modified perturbative 
approach. A gap in the spectrum which rises linearly has to be present in sectors excited by  other gluonic degrees of freedom such as the order 
parameter field $u(x_{\perp})$ defined in (\ref{I54}). Compactness of the Polyakov loop variables is not sufficient; variables like $u(x_{\perp})$ 
also have to display similar properties. Origin or mechanisms for making such variables compact have not been identified. In the present axial gauge 
representation it is tempting to connect such mechanisms to the presence of monopoles whose existence is intimately linked to the compactness of the 
Polyakov loop variables. As we have seen, due to the quantization of the electric flux,  the compactness leads to a  magnetic field dominated ground 
state and exhibits the dual Meissner effect. All these characteristic properties of the QCD vacuum are however only realized as far as the Polyakov 
loop variables are concerned. Compactness of the Polyakov loop variables on the other hand implies the presence of singular field configurations  
whenever the Polyakov loop variables reach the border of their domain. Condensation of the points where the Polyakov loop passes through the center 
of the group is made likely by the compactness of these variables. The processes which control the appearance of these singular fields are not 
understood. Unlike  't~Hooft-Polyakov monopoles of the Higgs model or  instantons in QCD, the singular field configurations emerging in the gauge 
fixing process, in general, are not solutions to the classical field equations and the associated action can be arbitrary small. Their appearance is 
therefore not suppressed by a necessarily large contribution to the action. On the other hand, the entropy generated by the quantum fluctuations will 
be reduced when monopoles are present. Only if the fluctuations satisfy certain boundary conditions (cf.\ Eq.(\ref{str})) the action will remain 
finite. This interplay between monopoles and fluctuations presents a  difficult theoretical problem. After decomposing a certain configuration  into 
singular and fluctuating components the action contains terms which are linear in the fluctuations. Furthermore a canonical description of this 
dynamics has not been derived. In our discussion of the monopole dynamics we will therefore adopt a phenomenological point of view and  on the basis 
of well understood properties of QCD we  will attempt  an indirect characterization of the dynamics.

In the first part we will present arguments concerning the frequency of occurrence of  singular field configurations in the partition function. 
To this end we make use of a relation between singular field configurations and instantons. 
 As is easily verified, the Polyakov loop of a single
instanton of size $\rho$ ($\rho \ll L$) is given by
\beqs 
P_{I}(x_{\perp})= e^{i\pi \mbox{\boldmath$\scriptstyle\tau $}{\bf x}_{\perp} 
/\sqrt{x_{\perp}^{2}+\rho^{2}}}
\eeqs 
Here and throughout this section we assume space-time to be Euclidean and use  vector notation for $x_{\perp}$ and corresponding quantities.
The above equation shows that, in axial gauge,  a single instanton contains a north ($P=1$) and south pole
singularity ($P=-1$) at its center and at infinity respectively. More generally it has be shown~\cite{JaLe98,QuRS99,FoTW99}
that the topological charge $\nu$ of a field configuration is given by the difference of
the net northern and southern charge
\beqs
  \nu = {\textstyle\frac12} \Bigl( \!\!
    \sum_{ i\atop P(x_{\perp i}) =1} \!\!\!\!\! m_i 
    \, -  \!\!\!\!\! \sum_{ i\atop P(x_{\perp i})=-1}\!\!\!\!\!  m_i 
  \Bigr) .
\eeqs
This relation implies, that field configurations which carry topological charge are  necessarily singular. This is plausible. In the axial gauge one 
removes almost completely one space time component of the gauge field. The remaining 3 components (together with some ``zero modes''), if continuous,
are not sufficient to generate a non-vanishing topological charge. Thus in order to cover topologically non-trivial sectors, the ensemble of gauge 
fixed field configurations must contain singular fields. Furthermore,  the action of singular fields with a topological charge, containing for 
instance   a north pole and a south antipole, cannot be smaller than the instanton action $ \frac{8\pi^2}{g^2}$.
The above relation can be used to estimate the density of monopoles $n_{M}$ on the basis of the density $n_{I}$ obtained in either the instanton 
liquid model~\cite{SCSH96} or from  lattice QCD~\cite{CGHN94}. In order to relate instanton and monopole
densities, we treat the instantons as independent; in this case the Polyakov
loop is given by a product over the Polyakov loops associated with single
instantons (or anti-instantons). For small instanton sizes $\rho \ll L
$, the number of instantons plus anti-instantons integrated over in the time
integral of the Polyakov loop for fixed ${\bf x}_{\perp}$ is of the order of $L
n_{\olabbr{I}}\rho^{3}$. On the average, the number of instantons and
anti-instantons will be the same and thus the phase accumulated in the time
integral will be given by the fluctuations and is thus expected to be of the 
order of $\pi \sqrt{L n_{\olabbr{I}}\rho^{3}}$. When changing the position
${\bf x}_{\perp}$ by an instanton size of $\rho$ we expect, in the absence of
correlations between the instantons, a different value of the phase which
however is of the same order of magnitude.  Thus, in this change of ${\bf x}_{\perp}$
the Polyakov loop will typically pass $\sqrt{L n_{\olabbr{I}}\rho^{3}}$
times through the center of the group which implies the following estimate for
the monopole density (counting both poles and antipoles) at low temperatures
\beq
\label{pm11}
  n_{\olabbr{M}} \propto \left(L  n_{\olabbr{I}}\rho\right)^{3/2}
  \quad , \quad \rho \ll L ,
\eeq
while for high temperatures, the integral involves at most one instanton and we
therefore expect
\beq
\label{pm12}
  n_{\olabbr{M}} \propto L  n_{\olabbr{I}} \quad , \quad \rho \ge L . 
\eeq
Surprisingly, the result (\ref{pm11}) implies an infinite monopole density at infinite extension or zero temperature. Although derived on the basis 
of a finite value of the instanton density, this result is to a large extent independent of the
particular model; the characteristic $L^{3/2}$ dependence is mostly a consequence
of the integration over $x_{3}$   which becomes ill-defined at infinite extension. It thus appears that irrespective of the temperature, axial
gauge monopoles occur with non-vanishing or even infinite density. 
 
Beyond generation of  monopoles via instantons, the system has the
additional option of producing one type (north or south) of poles
and corresponding antipoles only. No topological charge is associated with such singular fields and their occurrence is not limited by a bound on the 
action. Thus entropy  favors production of such configurations their suppression could only come through their coupling
to the quantum fluctuations. Since it appears that instantons may not be able
to account for confinement~\cite{CADG78} or, more precisely, for a realistic
value of the string constant~\cite{CGHN94}, this additional option could be relevant.
These entropy arguments also apply in the plasma phase.  For purely kinematical reasons a decrease in the monopole density must be expected as the 
estimates (\ref{pm11},\ref{pm12}) show. This decrease is counteracted by the enhanced probability to produce monopoles when,  with decreasing $L$, 
the Polyakov loop approaches more and more the center of the group, as has has been discussed in the previous section (cf.\ Fig.~\ref{Fig4}). One 
certainly has to expect a finite  density of singular fields also in the deconfined phase. In order for this to be compatible with the partially 
perturbative nature of the plasma phase and with dimensional reduction to QCD$_{2+1}$   poles and antipoles may have to  be 
strongly correlated with each other and to form effectively a gas of dipoles.
   
So far our qualitative arguments have dealt with the entropy associated
with a finite monopole density. We now address dynamical issues. Singular fields 
may contribute only little to the action. For this
reason we focus the discussion on the coupling of the quantum fluctuations $\delta A_{\mu}$ to the singular
fields $s_{\mu}$ (cf.\ Eq.(\ref{split})). 
We first consider  
 the coupling
of the charged quantum fluctuations to the Abelian monopole fields via the
4-gluon vertex,
\beqs
  \delta S = -g^{2}\int d^{4}x \sum_{i=1}^{3}
  \delta \mbox{\boldmath$\phi$}^{\dagger}(x) \cdot  \delta \mbox{\boldmath$\phi$}(x) \,\,
  {\bf a}^{\olabbr{s}}({\bf x}_{\perp})\cdot {\bf a}^{\olabbr{s}}({\bf x}_{\perp}) .
\eeqs
At long wave-lengths the quantum fluctuations thus acquire a mass $\delta m$ by
coupling to the Abelian magnetic field,
\beqs
  \delta m^{2} = g^{2}\frac{1}{V}\int_{V} d^{3}x \,
  {\bf a}^{\olabbr{s}}({\bf x}_{\perp}) \cdot {\bf a}^{\olabbr{s}}({\bf x}_{\perp}) .     
\eeqs
To compute this mass term we assume the singular field
${\bf a}^{\olabbr{s}}({\bf x}_{\perp})$ to be given by a superposition of the standard
monopole fields of Eq.(\ref{am18}),
\beqs
  {\bf a}^{\olabbr{s}}({\bf x}_{\perp})
  = \sum_{i=1}^{N} {\bf a}^{n_{i}} \left({\bf x}_{\perp}-{\bf x}_{\perp i}\right) 
\eeqs
with positions ${\bf x}_{\perp i}$ of the monopoles and  charges $n_{i} = \pm 1$ and
vanishing total magnetic charge,
\beqs
  \sum_{i=1}^{N} n_{i} = 0 .
\eeqs
In this way the functional integral over the Polyakov loop variables in
(\ref{pm0a}) has effectively been replaced by a summation over positions 
${\bf x}_{\perp i}$ and charges $n_{i}$ of these prescribed representative fields.
 Using standard identities from electrostatics, the mass term can be written as
\beq
\label{pm5}
  \delta m^{2} = \frac{1}{4 V}\int_{V} d^{3}x
  \sum_{i,j=1}^{N}\frac{n_{i}n_{j}}{|{\bf x}_{\perp}-{\bf x}_{\perp i}||{\bf x}_{\perp}-{\bf x}_{\perp j}|}
  = -\frac{\pi}{ V}\sum_{{\scriptstyle i,j=1\atop\scriptstyle i<j}}^{N}
  n_{i}n_{j}|{\bf x}_{\perp i}-{\bf x}_{\perp j}| .
\eeq
Similarly, neutral gauge field fluctuations acquire a mass $\delta m_3$ by coupling to the singular charged components,
\beqs
\delta m^{2}_{3} = g^{2}\frac{1}{V}\int_{V} d^{3}x \,
\mbox{\boldmath$\phi $}^{\olabbr{s}\,\dagger} ({\bf x}_{\perp}) \cdot
\mbox{\boldmath$\phi $}^{\olabbr{s}}({\bf x}_{\perp}) .     
\eeqs
Evaluation of this expression by using the charged partner (Eq.(\ref{za3})) of the Dirac monopole field yields
\beqs
\delta m^{2}_{3} = \delta m^{2} .     
\eeqs
Equation (\ref{pm5}) displays the resistance of the system to proliferate production of monopoles for entropy reasons. The mass  $\delta m^{2}$ 
decreases with decreasing distance between monopoles of equal charge. Coupling to quantum fluctuations thus  induces  attraction between monopoles of 
opposite charges  and repulsion between equal charge monopoles.  If strong enough the attraction  leads to monopole-antimonopole annihilation and 
ultimately to a vanishing $\delta m^{2}$.  Computation of the mass requires knowledge of the correlations between monopoles. For an uncorrelated gas, 
the mass vanishes. If, for instance, the dynamics  favors monopoles of opposite charge to be bound in dipoles and if the extension of the dipoles is 
much smaller than their separation, the expression (\ref{pm5}) can be simplified and yields a gluon mass
\beqs
\delta m^{2} = n_{\olabbr{M}} d
\eeqs
determined by  monopole density and  average dipole size $d$. If applied to the deconfined phase, it could provide a natural explanation for the 
appearance of a magnetic gluon mass. The temperature dependence of such a magnetic mass is determined by monopole density and dipole size. It is not 
obvious that such a mechanism of mass generation is appropriate for the confined phase. Here, non-perturbative mechanisms must be operative which 
distinguish between the center-parity even and odd states (cf.\ Eqs.(\ref{gap-},\ref{gap+})). Mass generation does not provide such a distinction, on 
the other hand,  boundary conditions imposed by the presence of monopoles  such as (\ref{str}) may do so. Fluctuations around singular fields have to 
satisfy these conditions  otherwise infinite action results. Given a finite monopole density, and assuming no particular correlations between the 
monopoles, this condition cannot be satisfied for long wave-length fluctuations, i.e.\ we expect fluctuations with wave number
\beqs
k \le k_{\mbox{min}} = n_{\olabbr{M}}^{1/3} 
\eeqs
to be dynamically suppressed. On the other hand, long wave length excitations associated with bilinears such as ${\bf \Phi}^{\dagger}(x)\cdot{\bf \Phi}(x)$ which generate $Z$-even states  are not necessarily affected by the boundary conditions  and may be associated with excitation energies which remain finite for infinite extensions.

\section{Conclusions}

We have discussed QCD in the framework of gauge fixed formulations. We have provided a detailed  analysis for both the Coulomb- and the axial gauge. The focus of our discussion has been on non-perturbative phenomena and we have  emphasized the role of compact variables and of singular field configurations which both arise as a result of fixing the gauge. The appearance of such peculiar configurations is well understood and reflects  obstructions in imposing these  gauge conditions. However, the gauge dependence of these exceptional configurations remains disturbing  and has been prohibitive for establishing the dynamical role of singular fields.  Existence of obstructions in gauge fixing may be the only property which is common to the various gauge fixed formulations of QCD. It is easily seen that  on a general level no statement concerning the importance  i.e.\ the weight in the generating functional of such configurations is possible. One may for instance  impose the Coulomb- or the axial gauge condition in the Georgi-Glashow model. These gauge conditions are independent of  matter fields and so are all the geometrical properties of the fields associated with the particular gauge fixing procedure. Nevertheless, depending on  the self-interaction of the Higgs particles, the system can be either in the Higgs or the confined phase. In the Higgs phase one does not expect these singular fields to be of any significance. The weight of the exceptional field configurations is thus determined by the dynamics. To clarify their relevance requires, dynamical studies beyond classification are required.

In Coulomb- and axial gauge representation of QCD, the role of the exceptional fields is very different. In Coulomb-gauge, Gribov horizons are given only implicitly and their presence affects all field configurations whose amplitudes exceed certain limits. In axial gauge only  a particular set of degrees of freedom, the Polyakov loop variables, are subject to the compactness requirements. This makes it much easier to implement compactness in the axial gauge, where the Gribov horizon is determined by geometrical properties. On the other hand, compactness of variables other than the Polyakov loops has to emerge dynamically within the axial gauge. Our analysis has provided  evidence that compact variables are the source of confinement-like phenomena.  This is a result of approximative calculations carried out in both gauges. Neglecting, apart from the compactness requirement,  all gluon self interactions the spectrum of the resulting Hamiltonians has been analyzed. In Coulomb-gauge, Gribov's approach amounts to replace the perturbative Hamiltonian of non-interacting massless gluons with an Hamiltonian of gluons with a momentum dependent effective mass which diverges in the infrared. In the axial gauge Hamiltonian, the compact Polyakov loop variables are treated like quantum mechanical degrees of freedom moving in an infinite square well. Common to both formulations is the suppression of elementary excitations at long wave-length and a deviation from the equipartition between electric and magnetic field energy. They both give rise to interaction energies of static charges which increase with distance. In axial gauge, a linear rise is obtained with the string constant given by the lattice strong coupling limit. In Coulomb-gauge a complete analysis has not been carried out; plausibility arguments also point to a linear increase. The two approaches  lead to distinctively different descriptions of the ground states corresponding to the compact variables. The approximation by a  momentum dependent gluon mass yields  electric field dominance,  the treatment of the compact variables as quantum mechanical degrees of freedom in an infinite square well gives rise to a magnetic vacuum.         

Improvement in the description  of the dynamics requires  gluon self interactions and possibly coupling to dynamical quarks to be taken into account. Such a calculation has been described for the axial gauge, results  in Coulomb-gauge do not exist. Confinement like properties get very much reduced in strength by coupling perturbatively the compact Polyakov loop variables to other gluonic variables. The linear rise in the  interaction energy of static charges persists. However the string constant decreases with the extension of the compact direction. The QCD scale generated by  the gluon loops does not at all affect the string constant. When dynamical quarks are coupled to the gauge fields, string-breaking occurs. It  is surprising that the minimal non-perturbative dynamics incorporated into the compactness of the Polyakov loops is sufficient to lead to this non-perturbative mechanism. Softening or loss of confinement must be expected to occur also in Coulomb-gauge.  The interaction energy receives loop  corrections from transverse gluons corresponding to the screening contribution to the running coupling constant. These corrections  are likely to lead to an interaction energy which remains finite for large distances since in Gribov's treatment the gluon  propagator actually vanishes in the infrared. Perturbative treatment of  gluonic self-interactions appears to be inappropriate for improving  the description of the confinement like phenomena linked to the presence of compact variables.   

Gauge fixed formulations are not only important for developing approximation schemes beyond standard perturbation theory, they also provide a useful framework for more phenomenologically oriented analyses of QCD. In particular axial gauge offers the possibility for a phenomenological approach to the Polyakov loop dynamics. These variables serve as order parameter fields to characterize the phases of QCD via the realization of the center symmetry. In the confined phase with the center symmetry realized, the states of the system can be classified (for SU($2$)) according to their behavior under center reflections. Hadronic states like glueballs have even center parity. Excitation energies of states of negative center parity, with  vanishing momentum in the compact direction, increase with the extension of the system. These two sectors of Yang-Mills theories are completely decoupled below the phase transition and merge into an interacting system above. Thus the phase transition cannot be understood as a reordering in the glueball sector, it involves a sector of the Hilbert space which is invisible above the critical extension or below the critical temperature. The Polyakov loops also characterize to some extent the high temperature phase. Analysis of thermodynamic and magnet stability limits the range of values the Polyakov loop can assume in the phase of broken center symmetry which in turn imply limits on the deviations of energy density and pressure from the Stefan-Boltzmann values.              

The most severe limitations in further developing gauge fixed formulation are related to the presence of singular field configurations. Unlike instantons or the 't Hooft-Polyakov monopole, singular field configurations in general do not satisfy the classical field equations. The action associated with such fields may be arbitrarily small.  Semiclassical treatments are not directly applicable. Singular fields and  fluctuations can mix. In general fluctuations around a singular field destroy the delicate balance of Abelian and non-Abelian contributions to the field strength resulting in an infinite action. There are no standard tools available to handle the difficult constraints imposed on the fluctuations by finiteness of the action. Essentially nothing is known about the treatment of singular fields in the canonical formalism. It is not obvious how to include them in an Ansatz for the QCD ground state wave-functional. Simple QED like structures for the wave-functional would exclude such fields by erroneously assigning to them an infinite energy. It seems unlikely that the issue of singular fields can be avoided. Smooth, but topologically non-trivial configurations must be in general expected to give rise to singular fields after gauge fixing. Once more one may approach these difficult dynamical issues also from a  phenomenological point of view. We have demonstrated this procedure by inferring from the high temperature limit of Yang-Mills theories, from results of lattice QCD or of models like the instanton liquid model some basic properties  of singular fields such as the their density, and its temperature dependence or their dynamical correlations. Despite some insights gained in this way we are still far away from a sufficiently complete phenomenological characterization of the dynamics of singular fields.   
\newpage
\bibliographystyle{unsrt}

\begin{thebibliography}{99}
\frenchspacing
\bibitem{SING78} I. M. Singer, {\em Comm. Math. Phys.} {\bf 60}, 7 (1978).
\bibitem{GRIB78} V. N. Gribov, {\em Nucl. Phys.} B {\bf 139}, 1 (1978).
\bibitem{MAND80} S. Mandelstam, {\em Phys. Rep.} {\bf 67}, 109 (1980).
\bibitem{THOO81} G. 't Hooft, {\em Nucl. Phys.} B {\bf 190}, 455 (1981).
\bibitem{BJOR} J.D. Bjorken, in: ``Lectures on Lepton Nucleon Scattering and Quantum Chromodynamics", W. Atwood {\em et al.}, Birkh\"{a}user (1982).
\bibitem{CHLE} N.H. Christ and T.D. Lee, {\em Phys. Rev.} D {\bf 22}, 939 (1980).
\bibitem{GESA} J.L. Gervais and B. Sakita, {\em Phys. Rev.} D {\bf 18}, 453 (1978).
\bibitem{CREU} M. Creutz, I.J. Muzinich, and T. N. Tudron, {\em Phys. Rev.} D {\bf 19}, 531 (1979). 
\bibitem{Georgi72} H. Georgi and S. Glashow, {\em Phys. Rev. Lett.} {\bf 28}, 1494 (1972). 
\bibitem{DREL} S.D. Drell, Asymptotic Freedom, ``A Festschrift for Maurice Goldhaber'', {\em Transact. of the New York Academy of Sciences} Series II, vol. {\bf 40}, 76 (1980).
\bibitem{Zwan82} D. Zwanziger, {\em Nucl. Phys.} B {\bf 209}, 336 (1982).
\bibitem{AnZw89} G. Dell'Antonio and D. Zwanziger, {\em Nucl. Phys.} B {\bf 326}, 333 (1989).
\bibitem{BAAL98} P. van Baal, in ``Continuous Advances in QCD 1998'', edt. by A. Smilga, World Scientific, 3  (1998). 
\bibitem{VSHA97} L. v. Smekal, A. Hauck and R. Alkofer, {\em Phys. Rev. Lett.} {\bf 79}, 3591 (1997).
\bibitem{ALVS00} R. Alkofer and L. v. Smekal, preprint hep-ph/0007355
\bibitem{BONN00} F. Bonnet et al., {\em Phys. Rev.} D {\bf 62}, 51501 (2000)
\bibitem{ALFF00} C. Alexandrou, Ph. de Forcrand, and E. Follana, preprints  hep-lat/0008012, hep-lat/0009003 
\bibitem{TDLE} T.D. Lee, ``Particle Physics and Introduction to Field Theory", Harwood Academic Publishers, Chur (1981).
\bibitem{SAV77} G.K. Savvidy, {\em Phys. Lett}. B {\bf 71}, 133 (1977).
\bibitem{NACH97} O. Nachtmann, in: Lectures on QCD, Applications, edt. by F. Lenz et al., Springer, 1 (1997). 
\bibitem{LeNT94} F. Lenz, H.W.L. Naus and M. Thies, {\em Ann. Phys.} {\bf 233}, 317 (1994).
\bibitem{JAKS97} O. Jahn, T. Kraus and M. Seeger, {\em Int. J. Mod. Phys.} A {\bf 12}, 4445 (1997).
\bibitem{McLerran81} L. McLerran and B. Svetitsky, {\em Phys. Lett.} B {\bf 98}, 195 (1981).
\bibitem{Svetitsky86} B. Svetitsky, {\em Phys. Rep.} {\bf 132}, 1 (1986).
\bibitem{Lenz98} F. Lenz and M. Thies, {\em Ann. Phys.} {\bf 268}, 308 (1998).
\bibitem{LeST95} F. Lenz, M. Shifman and M. Thies, {\em Phys. Rev.} D {\bf 51}, 7060 (1995).
\bibitem{LNRT00} F. Lenz, J.W. Negele, L. O'Raifeartaigh and M. Thies, hep-th/0004200, to appear in {\em Ann. Phys.}
\bibitem{SMIL94} A.V. Smilga, {\em Ann. Phys.} {\bf 234}, 1 (1994).
\bibitem{Kanayo} For a  review, see K. Kanayo, {\em Nucl. Phys.} B {\em (Proc. Suppl.)} {\bf 47}, 144 (1996).
\bibitem{Reisz} T. Reisz, {\em Z. Phys.} C {\bf 53}, 169 (1992).
\bibitem{DIRAC31} P.A.M. Dirac, {\em Proc. Roy. Soc. (London), Ser.} A {\bf 133}, 60 (1931).
\bibitem{GGHK94} B. Grossman, S. Gupta, U.M. Heller, and F. Karsch, {\em Nucl. Phys.} B {\bf 417}, 289 (1994). 
\bibitem{FIHK93} J. Fingberg, U. Heller, and F. Karsch, {\em Nucl. Phys.} B {\bf 392}, 493 (1993).
\bibitem{Schwinger} J. Schwinger, {\em Phys. Rev.} {\bf 130}, 402 (1963).
\bibitem{Weiss} N. Weiss, {\em Phys. Rev.} D {\bf 24}, 475 (1981).
\bibitem{LeMT95} F. Lenz, E.J. Moniz and M. Thies, {\em Ann. Phys.} {\bf 242}, 429 (1995).
\bibitem{KOSU} J. Kogut and L. Susskind, {\em Phys. Rev.} D , {\bf 11}, 395 (1975).
\bibitem{Uehling} E.A. Uehling, {\em Phys. Rev.} {\bf 48}, 55 (1935).
\bibitem{EKSM82} J. Engels, F. Karsch, H. Satz and I. Montvay, {\em Nucl. Phys}. B {\bf 205}, 545 (1982); {\em Phys. Lett}. B {\bf 101}, 89 (1981).
\bibitem{Engels} J. Engels, F. Karsch and K. Redlich, {\em Nucl. Phys}. B {\bf 435}, 295 (1995).
\bibitem{JaLe98} O. Jahn and F. Lenz, {\em Phys. Rev}. D {\bf 58}, 85006 (1998). 
\bibitem{QuRS99} M. Quandt, H. Reinhardt and A. Schafke, {\em Phys. Lett}. B {\bf 446}, 290 (1999). 
\bibitem{FoTW99} C. Ford, T. Tok and A. Wipf, {\em Phys. Lett}. B {\bf 456}, 155 (1999).
\bibitem{SCSH96} T. Sch\"afer and E.V. Shuryak, {\em Rev. Mod. Phys}. {\bf 70}, 323 (1998).
\bibitem{CADG78} C.G. Callan, R.F. Dashen and D.J. Gross, {\em Phys. Rev.} D {\bf 17}, 2717 (1978).
\bibitem{CGHN94} M.C. Chu, J.M. Grandy, S. Huang and J.W. Negele, {\em Phys. Rev.} D {\bf 49}, 6039 (1994).
\end{thebibliography}

\end{document}